\documentclass[10pt,aps,prd,superscriptaddress,nofootinbib,nobibnotes,longbibliography,floatfix,twocolumn]{revtex4-2}

\usepackage{bm}
\usepackage{mathtools,
amsmath,
amssymb,
amsfonts,
mathrsfs,
chngcntr,
multirow}

\let\cc\corresponds
\let\corresponds\relax
\usepackage{mathabx}
\let\corresponds\cc

\usepackage[utf8]{inputenc}
\usepackage[T1]{fontenc}

\usepackage{soul}

\usepackage[dvipsnames]{xcolor}
\usepackage[unicode]{hyperref}
\hypersetup{colorlinks=true, citecolor=MidnightBlue,
            linkcolor=MidnightBlue, urlcolor=MidnightBlue, linktocpage=true}
\usepackage[normalem]{ulem}

\usepackage{graphicx}

\usepackage{enumerate}

\usepackage{accents}
\newlength{\dhatheight}

\begin{document}
\title{A well-posed BSSN-type formulation for scalar-tensor theories of gravity with second-order field equations}

\author{Harry~L.~H.~Shum}
\address{Nottingham Centre of Gravity,
Nottingham NG7 2RD, United Kingdom}
\address{School of Mathematical Sciences, University of Nottingham,
University Park, Nottingham NG7 2RD, United Kingdom}

\author{Llibert Aresté Saló}
\address{Instituut voor Theoretische Fysica, KU Leuven. Celestijnenlaan 200D, B-3001 Leuven, Belgium. }
\address{Leuven Gravity Institute, KU Leuven. Celestijnenlaan 200D, B-3001 Leuven, Belgium. }

\author{Farid~Thaalba}
\address{Nottingham Centre of Gravity,
Nottingham NG7 2RD, United Kingdom}
\address{School of Mathematical Sciences, University of Nottingham,
University Park, Nottingham NG7 2RD, United Kingdom}
\address{SISSA, via Bonomea 265, 34136 Trieste, Italy \& INFN (Sez. Trieste), via Valerio 2, 34127 Trieste, Italy.}

\author{Miguel~Bezares}
\address{Nottingham Centre of Gravity,
Nottingham NG7 2RD, United Kingdom}
\address{School of Mathematical Sciences, University of Nottingham,
University Park, Nottingham NG7 2RD, United Kingdom}

\author{Thomas~P.~Sotiriou}
\address{Nottingham Centre of Gravity,
Nottingham NG7 2RD, United Kingdom}
\address{School of Mathematical Sciences, University of Nottingham,
University Park, Nottingham NG7 2RD, United Kingdom}
\address{School of Physics and Astronomy, University of Nottingham,
University Park, Nottingham NG7 2RD, United Kingdom}

\begin{abstract}
Recent developments in the modified harmonic and modified puncture gauges have opened new possibilities for performing stable numerical evolutions beyond General Relativity. In this work, we utilise techniques developed in the aforementioned formalisms to derive a BSSN-type formalism compatible with certain classes of modified gravity theories. As an intermediate step, we also derived modified versions of the Z4 and Z3 formalisms, thereby completing the connection between these formalisms beyond General Relativity. We then test the robustness of the new modified BSSN formalism by simulating the dynamics of black hole systems and benchmarking the results against the modified CCZ4 formulation. These developments enable the exploration of theories beyond General Relativity in many well-known Numerical Relativity codes that use different versions of the puncture gauge approach.
\end{abstract}

\maketitle

\section{Introduction}

Gravitational waves~\cite{LIGOScientific:2016aoc, LIGOScientific:2017vwq} have given us a new tool to probe the strong-field regime of gravity and test the validity of General Relativity (GR) under extreme conditions. Future detectors such as LISA~\cite{LISA:2017pwj}, the Einstein Telescope~\cite{Punturo:2010zz,ET:2019dnz}, and Cosmic Explorer~\cite{Reitze:2019iox} will enable us to detect gravitational waves from a larger variety of sources and with drastically increased precision. This motivates efforts to develop accurate waveform models of merging compact objects in and beyond GR theories.

To model the highly non-linear physics involved in the merger of similar-mass compact objects, one has to resort to numerical relativity (NR), where the field equations are solved numerically on a (super-)computer~\cite{Alcubierre,Baumgarte:2010, Bona:2009,Baumgarte:2021,Shibata:2015}. However, to obtain stable and physically meaningful numerical evolutions, the field equations must be written as a well-posed Cauchy problem~\cite{Hilditch:2013sba,Sarbach:2012pr}. Following Hadamard's criteria~\cite{Hadamard10030321135}, the well-posedness requirement demands that the solution of the system exists, is unique, and depends continuously on the initial data.  In particular, any small deviations in the initial data should lead to bounded changes in the final solution. In numerical implementations, the presence of discretisation errors will inevitably lead to small deviations from the exact initial data. Therefore, the well-posedness condition is crucial in ensuring that the growth of such errors remains bounded in subsequent evolutions.

Following the seminal work by Choquet–Bruhat that the Einstein equations admit a well-posed Cauchy problem~\cite{ChoquetBruhat:1952}, well-posedness is understood as a property of a particular formulation of a theory as an initial value problem; that is, a theory can be well-posed in a certain formalism and ill-posed in another. The NR community has developed several well-posed formulations of GR, including many $d+1$ formulations, which involve decomposing a $d+1$ dimensional spacetime into a collection of $d$ dimensional spatial hypersurfaces foliated over time. This approach includes the BSSNOK (Baumgarte-Shapiro-Shibata-Nakamura-Oohara-Kojima)~\cite{Shibata:1995we,Baumgarte:1998te,Nakamura:1987zz} (hereafter BSSN for brevity), the Z4~\cite{Bona:2003fj, Gundlach:2005eh}, the Z3~\cite{Bona:2003qn}, the Z4c~\cite{Bernuzzi:2009ex} and the conformal and covariant Z4 (CCZ4)~\cite{Alic:2011gg} formalisms. Another common way of evolving GR, in a well-posed system of equations, is by using generalised harmonic coordinates (GHC)~\cite{Pretorius:2004jg, Garfinkle:2001ni}, which evolves the full $d+1$ dimensional metric $g_{\mu \nu}$ using Einstein's equations. This approach was used to perform the first stable binary black hole (BH) evolution~\cite{PhysRevLett.95.121101}.

Unfortunately, none of these formulations lead to a well-posed problem out-of-the-box in many beyond-GR theories, as these can have a different partial differential equation (PDE) structure from GR. One can circumvent this problem by working perturbatively in the coupling constant that marks the deviation from GR~\cite{Benkel:2016rlz,Benkel:2016kcq,Okounkova:2017yby,Okounkova:2019dfo,Okounkova:2019zjf,Okounkova:2020rqw,Witek:2018dmd,Silva:2020omi,Elley:2022ept}, though this method is prone to a secular growth of errors and does not capture all non-linearities beyond GR~\cite{Reall:2021ebq,Allwright:2018rut,Okounkova:2019zjf,Okounkova:2020rqw}.

The well-posedness of Lovelock~\cite{Lovelock:1971yv} and Horndeski gravity~\cite{Horndeski:1974wa} --- the most general actions that lead to second order equations for a metric, and a metric and a scalar, respectively --- was considered in~\cite{Kovacs:2020pns,Kovacs:2020ywu}. It was proven that, in the weak coupling regime where the deviations from GR remain subdominant, these theories can be well-posed if one introduces a set of auxiliary metrics and imposes a new gauge fixing condition, known as the modified harmonic gauge (MHG). This formalism has already led to successful numerical evolutions beyond GR~\cite{East:2020hgw, East:2021bqk, East:2022rqi, Corman:2022xqg,Corman:2024vlk,Corman:2024cdr,Lara:2025kzj}.

Nevertheless, the use of the (modified) harmonic gauge in numerical simulations involving BHs necessitates the use of excision, where the numerical domain within the Apparent Horizon (AH) is suitably removed to avoid the evolution variables from reaching the BH singularity~\cite{Pretorius:2004jg,East:2020hgw}. This conceptually simple idea is, in fact, challenging to implement in evolution codes. To address this issue, the authors of~\cite{AresteSalo:2022hua} utilised the modified gauge and developed a puncture gauge formulation, the modified CCZ4 (mCCZ4), which applies to certain modified gravity theories. The puncture gauge formulation has the property that its coordinates are singularity-avoiding, thereby eliminating the need for excision. This formalism has been proven to be successful for the evolution of four-derivative scalar-tensor ($4\partial$ST) theories of gravity~\cite{AresteSalo:2023mmd,Doneva:2023oww,Doneva:2024ntw,AresteSalo:2025sxc,Hu:2025bkg} and recently~\cite{Corman:2025wun} showed its excellent agreement with the MHG formalism.  

These developments have significantly advanced our ability to extract predictions from beyond-GR theories in the deep non-linear regime. Nonetheless, there remains considerable room for further development. For instance, even though the CCZ4 formalism has been successfully extended to mCCZ4, the BSSN, Z4, and Z3 formalisms, each of which is known to be successful in different contexts, have yet to be extended beyond GR. In this paper, we address this gap and exploit the links among the BSSN, Z4, Z3, and CCZ4 formalisms to derive modified versions of the three former.  In particular, we focus on the modified BSSN (mBSSN) formalism, which requires the least modification to the original equations. Given the vast popularity of the BSSN formalism within the community, mBSSN would enable more research groups to perform numerical simulations beyond GR. 

Finally, it is worth noting that there are other approaches to address ill-posedness which can work in tandem with modified formulations. Indeed, the latter can become ill-posed for large values of the coupling constant and they are not readily applicable to theories with higher than second order derivatives. One such approach adopts techniques inspired by the Israel-Stewart formulation~\cite{Israel:1976tn} of dissipative hydrodynamics to cure well-posedness problems, commonly known as ``fixing-the-equations'' approach in modified gravity contexts~\cite{Cayuso:2017iqc,Cayuso:2020lca,Bezares:2021yek,Lara:2021piy,Gerhardinger:2022bcw,Franchini:2022ukz,Cayuso:2023xbc,Cayuso:2023dei,Gerhardinger:2024rza,Lara:2024rwa,Corman:2024cdr,Rubio:2024ryv}, which nevertheless involves prescribing ad-hoc evolution equations and additional time scales. Another promising avenue is to formulate beyond-GR theories in a well-posed manner by augmenting the theory with ``regularising'' terms generated by field redefinitions within the validity of the effective field theory(EFT)~\cite{Figueras:2024bba, Figueras:2025wtx}.

The paper is structured as follows: In Section~\ref{sec:evol_formalism}, we will first review the $d+1$ formalisms for GR in more detail, followed by a review of the modified gauge and mCCZ4.  We then present the derivation of the mBSSN formalism and study its well-posedness in GR and beyond.  In Section~\ref{sec:numerics}, we implement the mBSSN system numerically and study the evolution of two BH spacetimes, namely i) an isolated spinning BH and ii) a head-on merger of two BHs.  We then benchmark our results with simulations obtained from the mCCZ4 formalism.

We follow the conventions in Wald’s book~\cite{Wald:1984rg}. Greek letters $\mu,\nu,\ldots$ denote spacetime indices, and they run from $0$ to $d$; Latin letters $i,j,\ldots$ denote indices on the spatial hypersurfaces, and they run from $1$ to $d$. We set $c=G=1$.

\section{Evolution Formalisms}\label{sec:evol_formalism}

In this section, we provide a brief overview of the $d+1$ decomposition and discuss how different formalisms in Numerical Relativity can be built from this idea. To be concrete in our discussions, we will consider Einstein's equations and assume that the stress energy tensor $T^{\mu\nu}$ can describe ordinary matter, or be an effective stress-energy tensor that contains higher derivative terms and contributions from new fields, associated with deviations from GR. In the latter case, this effective stress-energy tensor could contain second derivatives of the metric, as we will discuss in due course.

\subsection{$d+1$ formalisms}\label{sec:d+1}
The $d+1$ decomposition of spacetime is a technique used in formulating GR as an initial value problem~\cite{smarryork,Choquet-Bruhat:2009xil}. The spacetime manifold is foliated into a collection of $d$-dimensional spatial hypersurfaces labelled by a ``time'' function~\cite{Arnowitt:1959ah,Wald:1984rg,Gourgoulhon:2007ue}. The geometric information is then encoded in the induced (spatial) metric $\gamma_{ij}$, and the extrinsic curvature $K_{ij}$. The induced metric describes the intrinsic geometry of a given hypersurface, and the extrinsic curvature determines how the hypersurface is embedded in the full spacetime. The extrinsic curvature can be related to the evolution of the spatial metric by $K_{ij}=-\frac{1}{2}\mathcal{L}_n \gamma_{ij}$, where $n^\mu$ is the (time-like) normal vector to a given hypersurface.

Using the $d+1$ decomposition, the line element of the full spacetime reads
\begin{equation}
    \mathrm{d}s^2=-\alpha^2 \mathrm{d}t^2 + \gamma_{ij}(\mathrm{d}x^i+\beta^i \mathrm{d}t)(\mathrm{d}x^j+\beta^j \mathrm{d}t)~,
\end{equation}
where $\alpha,~\beta^i$ are the lapse and shift vector, respectively. Using these coordinates, the normal to the $t=\text{constant}$ hypersurfaces can be written as $n^\mu = \frac{1}{\alpha}(\delta^\mu_0-\beta^i\delta^\mu_i)$. The spatial metric $\gamma_{ij}$ will be used to raise and lower the spatial indices.

After performing the $d+1$ decomposition, the field equations of GR reduce to evolution equations for $\gamma_{ij}$ and $K_{ij}$ as well as four constraint equations that must be satisfied on each hypersurface. Solving the Cauchy problem then amounts to determining how $\gamma_{ij}$ and $K_{ij}$ evolve over ``time'' $t$. However, the fields $\gamma_{ij}$ and $K_{ij}$ are not necessarily explicitly evolved. 
For example, in certain formalisms (such as the BSSN~\cite{Baumgarte:1998te,Shibata:1995we}), the preferred evolution variables are those obtained by performing a conformal decomposition of the fields. More precisely, the evolution variables (excluding the gauge variables $\alpha,\beta^i$) are taken to be
\begin{subequations}\label{def:conformal}
    \begin{align}
    \chi &= \det(\gamma_{ij})^{-\frac{1}{d}}~,\label{def:chi}\\
    \tilde{\gamma}_{ij} &= \chi \gamma_{ij}~,\label{def:gtilde}\\
    K &= \gamma^{ij}K_{ij}~,\label{def:K}\\
    \tilde{A}_{ij} &= \chi \big( K_{ij}-\frac{1}{d}\gamma_{ij}K \big)~,\label{def:Atilde} \\ 
    \tilde{\Gamma}^i &=\tilde{\gamma}^{kl}\tilde{\Gamma}^i_{kl}~,\label{def:Gammatilde}
\end{align}
\end{subequations}
where $\chi$ is the conformal factor, $\tilde{\gamma}_{ij}$ the conformal metric, $K$ the trace of the extrinsic curvature, $\tilde{A}_{ij}$ the conformal transformation of the tracefree part of the extrinsic curvature, and $\tilde{\Gamma}^i$ is related to the Christoffel symbols $\tilde{\Gamma}^i_{kl}$ associated with the conformal metric.

On the other hand, other formulations, such as the Z4~\cite{Bona:2003fj}, evolve $\gamma_{ij}$ and $K_{ij}$ directly, but introduce auxiliary fields as evolution variables to ensure well-posedness, in particular the $Z^{\mu}$ vector. The field equations are extended at the covariant level as
\begin{equation}\label{eqZ4}
    R_{\mu \nu} +\nabla_\mu Z_\nu +\nabla_\nu Z_\mu = \kappa \left(T_{\mu \nu}-\frac{1}{2}T g_{\mu \nu}\right)~,
\end{equation}
where $\kappa=8\pi$. In this formalism, the evolution variables (again omitting the gauge variables) are $\gamma_{ij},K_{ij},\Theta$, and $Z_i$, where $\Theta=n_\mu Z^\mu$ is the component of $Z_\mu$ that is normal to the hypersurface, and $Z_i$ denotes the spatial component of $Z_\mu$. The introduction of the $Z^\mu$ field promotes the constraint equations to dynamical degrees of freedom, such that satisfying the original constraints reduces to satisfying the algebraic constraints $Z^\mu=0$. Moreover, by adding damping terms for $Z^\mu$ in Eq.~\eqref{eqZ4}, one obtains a formalism that dynamically enforces exponential decays of the constraint violations~\cite{Gundlach:2005eh}. Finally, one can combine the two approaches of introducing auxiliary fields and performing a conformal decomposition of the evolution fields. The resultant formalism is known as CCZ4~\cite{Alic:2011gg}. 

Even though these formalisms produce a well-posed initial value problem for GR, they cannot be directly applied beyond GR for reasons mentioned in the Introduction. Nevertheless, it has been shown recently that by introducing a modified gauge, one could extend some GR techniques to certain classes of modified gravity theories. In the following section, we will review the concept of the modified gauge.

\

\subsection{Review of the modified gauge}
The modified gauge was first introduced in~\cite{Kovacs:2020pns,Kovacs:2020ywu}. Instead of constructing the gauge condition using the physical metric alone, the authors introduce two additional auxiliary metrics $\hat{g}^{\mu \nu}$ and $\tilde{g}^{\mu \nu}$ as part of the gauge fixing procedure. The auxiliary metrics $\hat{g}^{\mu \nu}$ and $\tilde{g}^{\mu \nu}$ can, in general, be freely specified, with the requirement that their null cones lie outside of the physical null cone and do not overlap with each other. In practice, this can be achieved by setting
\begin{subequations}
\begin{eqnarray}
    \tilde{g}^{\mu \nu}&=g^{\mu \nu} -b(x)n^\mu n^\nu~,\\
    \hat{g}^{\mu \nu}&=g^{\mu \nu} -a(x)n^\mu n^\nu~,
\end{eqnarray}
\end{subequations}
where $a(x), b(x)$ are functions satisfying either $0<a(x)<b(x)$, $0<b(x)<a(x)$, or $-1<a(x)<0<b(x)$. 
Using the auxiliary metrics, the authors generalise the standard harmonic gauge in GR to construct the MHG, where the gauge source function is given by
\begin{equation}
    H^\mu \equiv \tilde{g}^{\rho \sigma} \nabla_\rho \nabla_\sigma x^\mu= -\tilde{g}^{\rho \sigma}\Gamma^\mu_{\rho \sigma}\,.
\end{equation}
The MHG is fixed by demanding $H^\mu=0$.\footnote{Note that in the standard harmonic gauge, one chooses the source function to be $H^\mu \equiv g^{\rho \sigma} \nabla_\rho \nabla_\sigma x^\mu= -g^{\rho \sigma}\Gamma^\mu_{\rho \sigma}$.} In the MHG, the equations of motion (for GR) are 
\begin{eqnarray}
    R^{\mu \nu}-\frac{1}{2}R g^{\mu \nu}+2\hat{P}_\alpha^{\phantom{i} \beta \mu \nu}\nabla_\beta H^\alpha &= \kappa T^{\mu \nu},
    \label{eq:mhg_cov}
\end{eqnarray}
where $\hat{P}_\alpha^{\phantom{i} \beta \mu \nu}=\delta^{(\mu}_\alpha \hat{g}^{\nu )\beta} - \frac{1}{2}\delta^\beta_\alpha \hat{g}^{\mu \nu}$. In this formalism, the ``pure gauge'' modes are restricted to propagate along the null cone of $\tilde{g}^{\mu \nu}$, while the ``gauge-condition violating'' modes propagate along the null cone of $\hat{g}^{\mu \nu}$.  The ability to separate the propagation speeds of the ``physical'', ``pure gauge'', and ``gauge-condition violating'' modes is where the power of the modified gauge lies, as it has huge implications in extending the well-posedness results in GR to certain classes of modified theories of gravity.  We will return to this point in more detail in Section~\ref{sec:bGR_wellposedness}.

We note that, in the weakly-coupled regime, modified gravity theories belonging to the Lovelock and Horndeski family have been proven to be well-posed using the MHG~\cite{Kovacs:2020ywu}.

\subsection{Review of the modified CCZ4 formalism}
Recently, it has been shown that a CCZ4-type formalism can also be constructed for some beyond-GR theories of gravity by introducing auxiliary metrics into the system~\cite{AresteSalo:2022hua,AresteSalo:2023mmd}.
This formalism is known as the modified CCZ4 (or mCCZ4).  The covariant equations of motion of GR in the mCCZ4 formulation can be written as
\begin{widetext}
\begin{eqnarray}
    &R^{\mu \nu}-\frac{1}{2}R g^{\mu \nu}+2\hat{P}_\alpha^{\phantom{i} \beta \mu \nu}\nabla_\beta Z^\alpha - \kappa_1\left[2n^{(\mu}Z^{\nu)}+\left(\frac{d-3}{2+b(x)}+\frac{d-1}{2} \kappa_2 \right)n^\alpha Z_\alpha g^{\mu \nu}\right]= \kappa\,T^{\mu \nu}\,,
    \label{eq:mccz4_cov}
\end{eqnarray}
\end{widetext}    
where $\kappa_1$, $\kappa_2$ are the constraint damping constants that satisfy the bounds $\kappa_1>0, ~\kappa_2>-2/(2+b(x))$ derived in \cite{AresteSalo:2023mmd} (analogous bounds for GR formalisms can be found in~\cite{Alic:2011gg, Alic:2013xsa, Gundlach:2005eh, Pretorius:2004jg}).  
The gauge condition yields
\begin{equation}
    Z^\mu \equiv -\frac{1}{2}(\tilde{H}^\mu + \tilde{g}^{\rho \sigma}\Gamma^\mu_{\rho \sigma})=0~,
\end{equation}
where $\tilde{H}^\mu$ is a gauge source function that can be freely specified.\footnote{Note that when setting $\tilde{H}^{\mu}=0$ we recover the modified harmonic gauge formulation.} It can be shown that for suitable choices of $\tilde{H}^\mu$, one can derive the modified version of the 1+log slicing and Gamma-driver equations for the gauge~\cite{AresteSalo:2023mmd},
\begin{subequations}
\begin{align}
    \partial_{\perp}\alpha &= -\frac{2 \alpha}{1+a(x)}(K-2 \Theta)~, \label{eq:mod1+log}\\
    \partial_{\perp}\beta^i &= \frac{d}{2(d-1)}\frac{\hat{\Gamma}^i}{1+a(x)}-\frac{a(x)}{1+a(x)}\alpha D^i\alpha~, \label{eq:mod_Gamma_driver}
\end{align}
\end{subequations}
where $\partial_{\perp}=\partial_t-\beta^k\partial_k$. By performing a $d+1$ decomposition of~\eqref{eq:mccz4_cov} and a conformal decomposition of the evolution variables, one obtains the evolution equations of the mCCZ4 formalism (to be quoted in the following Section). As we depart from GR and consider modified gravity theories at weak coupling, the equations of motion of the theory can still be written in the form~\eqref{eq:mccz4_cov}, with the stress-energy tensor on the right-hand side (RHS) interpreted as the effective stress-energy tensor which contains the modified gravity corrections. In general, this will introduce time derivatives of our evolution fields on the RHS. To consistently move all time derivatives to the left-hand side (LHS), one has to perform a matrix inversion in the process of obtaining the equations of motion. We will revisit the details of this matrix inversion process in Appendix~\ref{app:inverseA}.

We note that mCCZ4 has been explicitly proven to be well-posed in the weak coupling regime for Einstein-Gauss-Bonnet gravity in $d=4$, and $4\partial$ST theories~\cite{AresteSalo:2023mmd}.

\section{Modified BSSN formalism}\label{section:modifed_bssn_formalism}
The primary focus of this work is to investigate whether a BSSN-type formalism can be developed for certain classes of modified theories of gravity, utilising the techniques established for the modified gauge. In both MHG and mCCZ4, one has to add a term associated with the projector $\hat{P}_\alpha^{\phantom{i} \beta \mu \nu}$ to the original equations of motion to make use of the modified gauge. It is not immediately obvious how one can adopt the same approach to the BSSN formalism.

Notwithstanding, in GR, the various $d+1$ formulations are known to be related. The CCZ4 is derived from a conformal decomposition of the Z4 system; the Z3 system can be derived from the Z4 following a ``symmetry-breaking'' procedure~\cite{Bona:2003qn}; and the BSSN appears as a special case of the Z3 system. Since the mCCZ4 formalism has proven to be successful for certain classes of beyond-GR theories, one might attempt to follow a similar line of reasoning to derive a BSSN-equivalent formalism. This is the approach we will follow in this work.

In this Section, we will outline our derivation of the mBSSN formalism from the mCCZ4 formalism.  We first decompose the stress-energy tensor in the usual manner:
\begin{subequations}\label{eqsTmunu}
\begin{eqnarray}
    \rho &=&n^\mu n^\nu T_{\mu \nu}\,,\label{rhomatter}\\
    S_i &=&-n^\mu \gamma_i^{\phantom i \nu}T_{\mu \nu}\,,\\
    S_{ij}&=&\gamma_i^{\phantom i \mu} \gamma_j^{\phantom j \nu}T_{\mu \nu}\,\label{Sij}.
\end{eqnarray}
\end{subequations}
The mCCZ4 equations for a $d+1$ spacetime, which will be our starting point for deriving the other formalisms, are:
\begin{widetext}
    \begin{subequations}\label{eqsccz4}
\begin{eqnarray}
\partial_{\perp}\tilde{\gamma}_{ij} &=& -2\alpha\tilde{A}_{ij}+2\tilde\gamma_{k(i}\partial_{j)}\beta^k-\tfrac{2}{d}\tilde{\gamma}_{ij}\partial_k\beta^k, \\
\partial_{\perp}\chi &=& \tfrac{2}{d}\chi\big(\alpha K - \partial_k\beta^k\big), \\
\partial_{\perp}K&=&-D^iD_i\alpha +\alpha\left[R+2\,D_iZ^i +K(K-2\Theta)\right] -d\,\kappa_1(1+\kappa_2)\,\alpha\,\Theta+\tfrac{\kappa\,\alpha}{d-1}\big[S-d\,\rho\big]\nonumber\\
&&\textstyle-\frac{d\,\alpha\,b(x)}{2(d-1)(1+b(x))}\Big[R-\tilde{A}_{ij}\tilde{A}^{ij}+\frac{d-1}{d}K^2 -(d-1)\kappa_1(2+\kappa_2)\,\Theta -2\,\kappa\,\rho\Big]\,,\\
\partial_{\perp}\Theta &=&{\textstyle\frac{\alpha}{2}}\big[R-\tilde A_{ij}\,\tilde A^{ij}+{\textstyle\frac{d-1}{d}}\,K^2 
+2\,D^iZ_i-2\,\Theta\,K\big]-Z_i\,D^i\alpha-\frac{\kappa_1}{2}\big(d+1+(d-1)\kappa_2\big)\,\alpha\,\Theta-\kappa\,\alpha\,\rho \nonumber\\
&&-\textstyle\frac{b(x)}{1+b(x)}\Big\{ \tfrac{\alpha}{2}\big( R-\tilde A_{ij}\,\tilde A^{ij}+{\textstyle\frac{d-1}{d}}\,K^2 \big)  -\frac{\kappa_1}{2}\big[\frac{d-3}{2+b(x)}+d+1+(d-1)\kappa_2\big]\,\alpha\,\Theta - \kappa\,\alpha\,\rho
\Big\}\,,\\
\partial_{\perp}\tilde{A}_{ij}&=&~\alpha[\tilde{A}_{ij}(K-2\Theta)-2\,\tilde{A}_{ik}\tilde{A}^k_{~j}]+2\tilde{A}_{k(i}\partial_{j)}\beta^k +\chi\left[\alpha\left(R_{ij} + 2D_{(i}Z_{j)}-\kappa\,S_{ij}\right)-D_iD_j\alpha\right]^{\text{TF}}\nonumber\\
&&- \tfrac{2}{d}(\partial_k\beta^k)\tilde{A}_{ij}\,,\\
\partial_{\perp}\hat\Gamma^i
&=&2\,\alpha\big(\tilde\Gamma^i_{\phantom i kl}\tilde A^{kl}-{\textstyle\frac{d-1}{d}}\tilde\gamma^{ij}\partial_jK-{\textstyle\frac{d}{2\,\chi}}\,\tilde A^{ij}\partial_j\chi\big) -2\,\tilde A^{ij}\partial_j\alpha-\hat\Gamma^j\partial_j\beta^i + {\textstyle\frac{2}{d}}\,\hat\Gamma^i\partial_j\beta^j + \tfrac{d-2}{d}\,\tilde\gamma^{ik}\partial_k\partial_j\beta^j + \tilde\gamma^{jk}\partial_j\partial_k\beta^i \nonumber\\
&&+2\,\alpha\,\tilde\gamma^{ij}\big(\partial_j\Theta- {\textstyle\frac{1}{\alpha}}\,\Theta\,\partial_j\alpha - \tfrac{2}{d}\,K\,Z_j\big) -2\,\kappa_1\,\alpha\,\tilde\gamma^{ij}Z_j\,-2\,\kappa\,\alpha\,\tilde\gamma^{ij}S_j \nonumber\\
&& -\textstyle\frac{2\alpha\,b(x)}{1+b(x)}\Big[
\tilde D_j\tilde A^{ij}-\big(\tfrac{d-1}{d}\big)\tilde\gamma^{ij}\partial_jK-\tfrac{d}{2\,\chi}\tilde A^{ij}\partial_j\chi+\tilde\gamma^{ij}\big(\partial_j\Theta-\tfrac{1}{d}\,K\,Z_j\big) - \tilde A^{ij}Z_j -\kappa_1\,\tilde\gamma^{ij}\,Z_j-\kappa\,\tilde\gamma^{ij}S_j
\Big]\,,
\end{eqnarray}
\end{subequations}
\end{widetext}
where $S=\gamma^{ij}S_{ij}$, and $\hat{\Gamma}^i$ is related to the previously defined $\tilde{\Gamma}^i$ by
\begin{equation}
    \hat{\Gamma}^i = \tilde{\Gamma}^i + 2\tilde{\gamma}^{ij}Z_j~,
    \label{def:Gammahat}
\end{equation}
and $\left[\dots\right]^{\text{TF}}$ refers to the trace-free part of the expression. To derive the modified Z4 formalism from the mCCZ4 formalism, we ``undo'' the conformal transformation by substituting the definition of the conformal variables~\eqref{def:conformal}
and~\eqref{def:Gammahat} into the mCCZ4 equations \cite{AresteSalo:2023mmd}. This allows us to rewrite the evolution equations of the system in terms of the Z4 variables, which provides the equations for the modified Z4 (mZ4) formalism:
\begin{widetext}
\begin{subequations}\label{eqsz4}
\begin{eqnarray}
\partial_{\perp}\gamma_{ij} &=& -2\alpha K_{ij}+2\gamma_{k(i}\partial_{j)}\beta^k, \\
\partial_{\perp}K_{ij}&=&-D_iD_j\alpha +\alpha\left[R_{ij}+2\,D_{(i}Z_{j)} +K_{ij}(K-2\Theta) -2\,K_{ik}K^k_{~j}\right]
-\kappa\,\alpha\big[S_{ij}-\frac{\gamma_{ij}}{d-1}\,(S-\rho)\big]\textstyle-\gamma_{ij}\kappa_1(1+\kappa_2)\,\alpha\,\Theta \nonumber\\
&&-\frac{\gamma_{ij}}{2(d-1)}\frac{\alpha\,b(x)}{1+b(x)}\Big[R-K_{kl}K^{kl}+K^2 
-(d-1)\kappa_1(2+\kappa_2)\,\Theta -2\,\kappa\,\rho\Big]+2K_{k(i}\partial_{j)}\beta^k\,,\\
\partial_{\perp}\Theta &=&{\textstyle\frac{\alpha}{2}}\big[R-K_{ij}\,K^{ij}+\,K(K-2\Theta) 
+2\,D^iZ_i]-Z_i\,D^i\alpha 
-\frac{\kappa_1}{2}\big(d+1+(d-1)\kappa_2\big)\,\alpha\,\Theta-\kappa\,\alpha\,\rho \nonumber\\
&&-\textstyle\frac{b(x)}{1+b(x)}\Big\{ \tfrac{\alpha}{2}\big( R-K_{ij}\,K^{ij}+\,K^2 \big) 
-\frac{\kappa_1}{2}\big[\frac{d-3}{2+b(x)}+d+1+(d-1)\kappa_2\big]\,\alpha\,\Theta - \kappa\,\alpha\,\rho
\Big\}\,,\\
\partial_{\perp}Z_i
&=&\alpha\big(D_j(\,K_{i}^{~j}-\delta_i^jK)+D_i\Theta-2K_i^{~j}Z_j\big) -\Theta\,D_i\alpha  -\kappa \alpha S_i 
- \kappa_1 \alpha Z_i+Z_j \partial_i \beta^j\nonumber\\&&
-\textstyle\frac{\alpha\,b(x)}{1+b(x)}\Big[
D_j(\,K_{i}^{~j}-\delta_i^jK) + D_i\Theta - K_i^{~j}Z_j -\kappa_1\,Z_i-\kappa\,S_i
\Big]~.
\end{eqnarray}
\end{subequations}
\end{widetext}
Using the mZ4 equations, we can straightforwardly derive the equations for the ``modified Z3'' (mZ3) formalism by following the same ``symmetry-breaking'' procedure introduced in GR \cite{Bona:2003qn}:  first, we redefine the extrinsic curvature as $K_{ij} \to \bar{K}_{ij} + \frac{n}{2}\Theta\gamma_{ij}$, then we set $\Theta=0$ to obtain a one-parameter family of evolution systems with parameter $n$. We may then simplify our notation by renaming $\bar{K}_{ij}$ back to $K_{ij}$. Performing this operation on the mZ4 equations produces the mZ3 formalism:
\begin{widetext}
\begin{subequations}\label{eqsz3}
\begin{eqnarray}
\partial_{\perp}\gamma_{ij} &=& -2\alpha K_{ij}+2\gamma_{k(i}\partial_{j)}\beta^k, \\
\partial_{\perp}K_{ij}&=&-D_iD_j\alpha +\alpha\left[R_{ij}+2\,D_{(i}Z_{j)} +K_{ij}K -2\,K_{ik}K^k_{~j}\right] 
-\kappa\,\alpha\big[S_{ij}-\frac{\gamma_{ij}}{d-1}\,(S-\rho)\big]+2K_{k(i}\partial_{j)}\beta^k  \nonumber\\&& \textstyle-\frac{\gamma_{ij}}{2(d-1)}\frac{\alpha\,b(x)}{1+b(x)}\Big[R-K_{kl}K^{kl}+K^2  -2\,\kappa\,\rho\Big]
-{\textstyle\frac{n\alpha}{4}}\gamma_{ij}\big[R-K_{lm}\,K^{lm}+\,K^2 
+2\,D^lZ_l] \nonumber\\&&+\textstyle\frac{n}{2}\frac{b(x)}{1+b(x)}\gamma_{ij}\Big\{ \tfrac{\alpha}{2}\big( R-K_{lm}\,K^{lm}+\,K^2 \big)  - \kappa\,\alpha\,\rho
\Big\}
+\frac{n}{2}\gamma_{ij}Z_l\,D^l\alpha+\frac{n}{2}\kappa\,\alpha\,\rho\,\gamma_{ij} \,,\\
\partial_{\perp}Z_i
&=&\alpha\big(D_j(\,K_{i}^{~j}-\delta_i^jK)-2K_i^{~j}Z_j\big)  -\kappa \alpha S_i - \kappa_1 \alpha Z_i
+Z_j \partial_i \beta^j\nonumber\\&&
-\textstyle\frac{\alpha\,b(x)}{1+b(x)}\Big[
D_j(\,K_{i}^{~j}-\delta_i^jK)- K_i^{~j}Z_j  -\kappa_1\,Z_i-\kappa\,S_i
\Big]~.
\end{eqnarray}
\end{subequations}
\end{widetext}
Lastly, we need to connect the mZ3 to the mBSSN formalism. In GR, it has been shown that the Z3 system is quasi-equivalent (equivalent in principal parts) 
to the BSSN system when $n=\frac{4}{d}$\,\cite{Bona:2003qn}.\footnote{We have generalised $n=\frac{4}{3}$ (for $3+1$ spacetimes~\cite{Bona:2003qn}) to $n=\frac{4}{d}$.} 
The authors noted that the ``BSSN'' system obtained in this way differs from the original BSSN system because it includes lower-order terms proportional to $Z^i$.  Nevertheless, by setting $Z_\mu = 0$, the standard BSSN equations \cite{Baumgarte:1998te} can be fully recovered.  

In short, the BSSN system can be obtained from the Z3 system by rewriting the equations in terms of the conformal variables~\eqref{def:chi},~\eqref{def:gtilde},~\eqref{def:K}, and~\eqref{def:Atilde} and defining $\tilde{\Gamma}_i=-\tilde{\gamma}_{ik}\partial_j\tilde{\gamma}^{kj}+2Z_i$ (note that this is the same definition as $\hat{\Gamma}^i$, but we will denote the LHS as $\tilde{\Gamma}^i$ to follow the standard notation in BSSN). After expressing the mZ3 equations in terms of the new fields, we set $Z_i \rightarrow 0$ and fix $n=\frac{4}{d}$. 
The outcome of these operations gives the mBSSN system:
\begin{widetext}
\begin{subequations}\label{eqsbssn}
\begin{eqnarray}
    \partial_{\perp}\tilde{\gamma}_{ij} &=& -2\alpha\tilde{A}_{ij}+2\tilde\gamma_{k(i}\partial_{j)}\beta^k-\tfrac{2}{d}\tilde{\gamma}_{ij}\partial_k\beta^k, \\
\partial_{\perp}\chi &=& \tfrac{2}{d}\chi\big(\alpha K - \partial_k\beta^k\big), \\
\partial_{\perp}K&=&-D^iD_i\alpha +\alpha\left[ \tilde{A}_{lm}\tilde{A}^{lm}+\frac{1}{d}K^2\right] +\tfrac{\kappa\,\alpha}{d-1}\big[S+(d-2)\,\rho\big]\nonumber\\&&
\textstyle\qquad+\frac{\alpha\,b(x)\,(d-2)}{2(d-1)(1+b(x))}\Big[R-\tilde{A}_{ij}\tilde{A}^{ij}+\frac{d-1}{d}K^2 -2\,\kappa\,\rho\Big]\,,\\
\partial_{\perp}\tilde{A}_{ij}&=&~ \chi\left[\alpha\left(R_{ij}-\kappa\,S_{ij}\right)-D_iD_j\alpha\right]^{\text{TF}} + \alpha \left(K \tilde{A}_{ij} - 2 \tilde{A}_{ik}\tilde{A}^k_{~j}\right) 
+2\tilde{A}_{k(i}\partial_{j)}\beta^k - \tfrac{2}{d}(\partial_k\beta^k)\tilde{A}_{ij}\,,\\
\partial_{\perp}\tilde{\Gamma}^i
&=&-2\,\tilde A^{ij}\partial_j\alpha-\tilde{\Gamma}^j\partial_j\beta^i + {\textstyle\frac{2}{d}}\,\tilde{\Gamma}^i\partial_j\beta^j + \tfrac{d-2}{d}\,\tilde\gamma^{ik}\partial_k\partial_j\beta^j 
+ \tilde\gamma^{jk}\partial_j\partial_k\beta^i\nonumber\\
&& +\textstyle 2\alpha\Big[
\tilde{\Gamma}^i_{jk}\tilde{A}^{jk}-\big(\tfrac{d-1}{d}\big)\tilde\gamma^{ij}\partial_jK-\tfrac{d}{2\,\chi}\tilde A^{ij}\partial_j\chi-\kappa\,\tilde\gamma^{ij}S_j
\Big]\nonumber\\&& -\textstyle\frac{2\alpha b(x)}{1+b(x)}\Big[
\tilde D_j\tilde A^{ij}-\big(\tfrac{d-1}{d}\big)\tilde\gamma^{ij}\partial_jK-\tfrac{d}{2\,\chi}\tilde A^{ij}\partial_j\chi-\kappa\,\tilde\gamma^{ij}S_j
\Big]\,.
\end{eqnarray}
\end{subequations}
\end{widetext}

Note that the modified gauge enters only through corrections to the evolution equations of $K$ and $\tilde{\Gamma}^i$:  it contributes a term proportional to the GR Hamiltonian constraint $\mathcal H$ in the evolution equation of $K$; and a term related to the GR momentum constraint ${\mathcal M}^i$ in the evolution equation of $\tilde{\Gamma}^i$, where
\begin{subequations}
    \begin{eqnarray}
        \hspace{-1cm}{\mathcal H}&=&R-\tilde{A}_{ij}\tilde{A}^{ij}+\frac{d-1}{d}K^2 -2\,\kappa\,\rho\,,\label{Ham_eq}\\
        \hspace{-1cm}{\mathcal M}^i&=&D_j\tilde A^{ij}-\big(\tfrac{d-1}{d}\big)\tilde\gamma^{ij}\partial_jK-\tfrac{d}{2\,\chi}\tilde A^{ij}\partial_j\chi-\kappa\,\tilde\gamma^{ij}S_j\,.
    \end{eqnarray}
\end{subequations} 
In the standard BSSN formulation of GR, one has to add a momentum constraint to the evolution equation of $\tilde{\Gamma}^i$ by hand during the derivation process.  This allows the cancellation of the term $\partial_j \tilde{A}^{ij}$ in the equation, which is known to cause numerical instabilities~\cite{Alcubierre}.  In our derivation of BSSN, this momentum constraint (along with the momentum constraint proportional to the modified gauge) is provided by the equation $\partial_{\perp}Z_i$ of the mZ3 formalism.  However, to use this equation in our derivation of mBSSN, we emphasise that it is important to keep the variable $Z^i$ non-zero throughout the derivation, and set $Z^i=0$ only in the final step.  Otherwise, we will not be able to remove the term $\partial_j \tilde{A}^{ij}$ from the equation, and the modified gauge will not be consistently incorporated into the evolution equation of $\tilde{\Gamma}^i$ in mBSSN.

\subsection{Well-posedness in GR}\label{sec:well-posednessGR}
In this section, we establish the well-posedness of the mBSSN formalism in GR.  Well-posedness implies that, for appropriate initial data, a unique solution to the field equations exists and depends continuously on the initial data. This final property ensures that the difference between two solutions starting from slightly different initial data will not grow without bound.

Mathematically, whether a formalism is well-posed or not can be determined through a hyperbolicity analysis~\cite{Sarbach:2012pr}. Roughly speaking, this amounts to studying the high-frequency behaviour of the PDE system. A formalism is locally well-posed if it is strongly hyperbolic, which requires the principal part (the highest derivative terms) of the equations to have real eigenvalues with a complete set of eigenvectors~\cite{Hilditch:2013sba}. For GR, it is sufficient to analyse the case of a vanishing $T^{\mu\nu}$, assuming that $T^{\mu\nu}$ does not contain second derivatives of the metric or the matter field. This assumption is true for ordinary matter that couples minimally to the metric and it implies that $T^{\mu\nu}$ does not contribute to the principal part of the PDE system.

To perform the analysis, we must complete the system by selecting a suitable gauge. Here, we will consider the modified $1+\log$ slicing and Gamma-driver equation~\eqref{eq:mod1+log},~\eqref{eq:mod_Gamma_driver} proposed in~\cite{AresteSalo:2023mmd} with suitable adaptations to the BSSN system, i.e., $\Theta \to 0,~\hat{\Gamma}^i \to \tilde{\Gamma}^i$. We will closely follow the procedure described in~\cite{Brown:2011qg, AresteSalo:2023mmd} when analysing the hyperbolicity of the system. First, we write down the principal part of the system,\footnote{For this purpose, terms such as $\tilde{A}_{ij}, K$ can be viewed as related to the ``velocity'' of the spatial metric and hence are one order above $\gamma_{ij}$.}
\begin{subequations}
\label{eqsbssnpp}
\begin{eqnarray}
    \hspace{-0.5cm}\partial_{\perp}\tilde{\gamma}_{ij} &\cong& -2\alpha\tilde{A}_{ij}+2\tilde\gamma_{k(i}\partial_{j)}\beta^k-\tfrac{2}{d}\tilde{\gamma}_{ij}\partial_k\beta^k\,, \\
\hspace{-0.5cm}\partial_{\perp}\chi &\cong& \tfrac{2}{d}\chi\big(\alpha K - \partial_k\beta^k\big)\,, \\
\hspace{-0.5cm} \partial_{\perp}K&\cong&-\gamma^{ij} \partial_i \partial_j\alpha+\frac{\alpha\,b(x)\,(d-2)}{2(d-1)(1+b(x))}(R)_\text{p.p.}\,,\\
\hspace{-0.5cm}\partial_{\perp}\tilde{A}_{ij}&\cong&~ \chi\left[\alpha (R_{ij})_\text{p.p.}-\partial_i \partial_j\alpha\right]^{\text{TF}}\,,\\
\hspace{-0.5cm}\partial_{\perp}\tilde{\Gamma}^i
&\cong&\tfrac{d-2}{d}\,\tilde\gamma^{ik}\partial_k\partial_j\beta^j + \tilde\gamma^{jk}\partial_j\partial_k\beta^i -2\alpha\partial_j\tilde{A}^{ij} \nonumber\\&& +\textstyle\frac{2\alpha}{1+b(x)}\Big[
\partial_j\tilde A^{ij}-\big(\tfrac{d-1}{d}\big)\tilde\gamma^{ij}\partial_jK
\Big]\,,\\
\hspace{-0.5cm}\partial_{\perp}\alpha &\cong& -\frac{2 \alpha}{1+a(x)}K\,, \\
\hspace{-0.5cm}\partial_{\perp}\beta^i &\cong& \frac{d}{2(d-1)}\frac{\tilde{\Gamma}^i}{1+a(x)}-\frac{a(x)}{1+a(x)}\alpha D^i\alpha\,,
\end{eqnarray}
\end{subequations}
where $(R)_\text{p.p.}$ and $(R_{ij})_\text{p.p.}$ denote the principal part of $R$ and $R_{ij}$ respectively.
From the principal part of the system, we can define the principal symbol as follows. We first label the fields that contribute to the highest order with a ``hat'' (e.g., $-2\alpha\tilde{A}_{ij} \to -2\alpha\hat{\tilde{A}}_{ij},~2\tilde\gamma_{k(i}\partial_{j)}\beta^k \to 2\tilde\gamma_{k(i}\partial_{j)}\hat{\beta}^k$). We then perform the replacement $\partial_{\mu}\to\xi_{\mu}\equiv(\mu,n_i)$, where $\xi_{\mu}$ is an arbitrary characteristic covector, with $n_i$ being its spatial part (satisfying $\gamma^{ij}n_in_j=1$),\footnote{This $n^i$ is not to be confused with the normal vector of a hypersurface introduced earlier.}
which gives us the principal symbol:
\begin{subequations}\label{eqsbssnps}
\begin{eqnarray}
    \check{\mu}\hat{\tilde{\gamma}}_{ij} &=& -2\alpha\hat{\tilde{A}}_{ij}+2\tilde\gamma_{k(i}n_{j)}\hat{\beta}^k-\tfrac{2}{d}\tilde{\gamma}_{ij}n_k\hat{\beta}^k\,, \\
\check{\mu}\hat{\chi} &=& \tfrac{2}{d}\chi\big(\alpha \hat{K} - n_k\hat{\beta}^k\big)\,, \\
\check{\mu}\hat{K}&=&-\hat{\alpha}+\frac{\alpha\,b(x)\,(d-2)}{2(d-1)(1+b(x))}\big( (d-1) \chi^{-1} \hat{\chi} \nonumber\\&&\hspace{3.5cm}+ \chi^{-1} n^in^j\hat{\tilde{\gamma}}_{ij} \big)\,,\\
\check{\mu}\hat{\tilde{A}}_{ij}&=& \chi \alpha \Big[  \frac{1}{2} \big(\tilde{\gamma}_{jm}n_i\hat{\tilde{\Gamma}}^m-\chi^{-1}\hat{\tilde{\gamma}}_{ij} +\chi^{-2} \tilde{\gamma}_{ij}\hat{\chi} \nonumber\\&&\hspace{0.5cm}+\chi^{-1}(d-2)n_in_j\hat{\chi}+\tilde{\gamma}_{im}n_j\hat{\tilde{\Gamma}}^m \big)\nonumber \\&&\hspace{0.5cm} -\big( (d-1)\chi^{-1} \hat{\chi} +\chi n_k\hat{\tilde{\Gamma}}^{k} \big) \frac{\gamma_{ij}}{d}\Big]  \nonumber \\&&-\chi \big( n_i n_j - \frac{1}{d} \gamma_{ij} \big) \hat{\alpha}
\,,\\
\check{\mu}\hat{\tilde{\Gamma}}^i
&=&\tfrac{d-2}{d}\,\tilde\gamma^{ik}n_kn_j \hat{\beta}^j + \tilde\gamma^{jk}n_jn_k\hat{\beta}^i -2\alpha n_j\hat{\tilde{A}}^{ij} \nonumber\\&& +\textstyle\frac{2\alpha}{1+b(x)}\Big[
n_j\hat{\tilde A}^{ij}-\big(\tfrac{d-1}{d}\big)\tilde\gamma^{ij}n_j \hat{K}
\Big]\,,\\
\check{\mu}\hat{\alpha} &=& -\frac{2 \alpha}{1+a(x)}\hat{K}\,, \\
\check{\mu}\hat{\beta}^i &=& \frac{d}{2(d-1)}\frac{\hat{\tilde{\Gamma}}^i}{1+a(x)}-\frac{a(x)}{1+a(x)}\alpha \gamma^{ij} n_j \hat{\alpha}\,,
\end{eqnarray}
\end{subequations}
where $\check{\mu}= \mu - \beta^i n_i$, and we have substituted the relevant expressions for $(R)_\text{p.p.}$ and $(R_{ij})_\text{p.p.}$ in our equations. In the following, we will use $\hat{R},\hat{R}_{ij}$ to denote the contributions to the principal symbol (instead of the principal part) coming from $R,R_{ij}$, respectively.

Note that there are different equivalent ways to write $R,R_{ij}$ (hence also $\hat{R},\hat{R}_{ij}$) because the field variables are not independent of each other (namely $\tilde{\Gamma}_i=-\tilde{\gamma}_{ik}\partial_j\tilde{\gamma}^{kj}$). For example, focusing on the contributions to the principal symbol alone, there are two different ways in which the trace-free part of $R_{ij}$ ($\hat{R}_{ij}^{TF}$) can be expressed in terms of our dynamical fields:
\begin{subequations}
 \begin{eqnarray}
    \hspace{-0.5cm}\hat{R}_{ij}^{TF}(\tilde{\gamma}) &=& \frac{\chi ^{-2}}{2} \big( \chi ^{-1} \hat{\chi} \tilde{\gamma}_{ij}-\hat{\tilde{\gamma}}_{ij}-\chi ^{-1} \hat{\chi} n_i \tilde{\gamma}_{\perp j}\nonumber\\&&+n_i \hat{\tilde{\gamma}}_{\perp j} \nonumber +n_i n_j \hat{\chi}d- n_in_j \chi \tilde{\gamma}^{km}\nonumber \\&&\hat{\tilde{\gamma}}_{km}-\chi^{-1}n_j \hat{\chi} \tilde{\gamma}_{\perp i}+n_j \hat{\tilde{\gamma}}_{\perp i} \big)\nonumber \\&&-\big( (d-1)\chi^{-1}\hat{\chi} + \chi^{-1}n^kn^m\hat{\tilde{\gamma}}_{km} \big)\frac{\gamma_{ij}}{d}\,,\\
   \hspace{-0.5cm} \hat{R}_{ij}^{TF} (\tilde{\Gamma})&=&  \frac{1}{2} \big(\tilde{\gamma}_{jm}n_i\hat{\tilde{\Gamma}}^m-\chi^{-1}\hat{\tilde{\gamma}}_{ij}+\chi^{-2} \tilde{\gamma}_{ij}\hat{\chi} \nonumber\\&&+\chi^{-1}(d-2)n_in_j\hat{\chi}+\tilde{\gamma}_{im}n_j\hat{\tilde{\Gamma}}^m \big) \nonumber \\&&-\big( (d-1)\chi^{-1} \hat{\chi} +\chi \hat{\tilde{\Gamma}}^{\perp} \big) \frac{\gamma_{ij}}{d}~,\label{eqn:R_TF_Gamma}
\end{eqnarray}
\end{subequations}
where we have denoted the projection of an index along $n^i$ by ``$\perp$''.  Similarly for $\hat{R}$, we can write
\begin{subequations}
\begin{align}
    \hat{R}(\tilde{\gamma}) &= (d-1) \chi^{-1} \hat{\chi} + \chi^{-1} \hat{\tilde{\gamma}}_{\perp \perp} \label{eqn:R_gamma}~,\\
    \hat{R}(\tilde{\Gamma}) &= (d-1) \chi^{-1} \hat{\chi} + \chi \hat{\tilde{\Gamma}}^\perp~.
\end{align}
\end{subequations}
The relation between $\tilde{\gamma}_{ij}$ and $\tilde{\Gamma}_i$ provides multiple ways of expressing the Ricci tensor or scalar in terms of our evolution variables. For our purposes, we will use the expression~\eqref{eqn:R_gamma} for the Ricci scalar to compute the modified gauge terms, while using~\eqref{eqn:R_TF_Gamma} for the rest of the terms to evaluate our principal symbol. This allows us to parallel the hyperbolicity analysis of mCCZ4, and as we will see, also results in the mBSSN reproducing the same eigenvalues as the mCCZ4 system. Choosing a different way of writing our principal symbol will affect the eigenvalues obtained from the system and the hyperbolicity of the equations.\footnote{This manifests most significantly in the equation of $\check{\mu}\hat{\tilde{A}}_{\perp A}$ (see the discussion on the vector block of the equations). If we write $\hat{R}_{ij}$ in a way such that it only contains projections of $\tilde{\gamma}_{ij}$ but with no $\tilde{\Gamma}^i$ terms, then the RHS of $\check{\mu}\hat{\tilde{A}}_{\perp A}$ will vanish. This implies a loss of hyperbolicity in the system.}
It is convenient at this point to decompose our system into different blocks and analyse the hyperbolicity of each block separately.  This can be done by introducing $d-1$ spatial unit vectors $e^i_A$ (with $A=1,...,d-1$), which form a tetrad in a way such that $n_i e^i_A=0$ and $e^i_A e^j_B \gamma_{ij} =~\delta_{AB}$. 

The principal symbol may then be decomposed into scalar, vector, and traceless tensor blocks by considering different projections with respect to the vectors $e^i_A$. The traceless tensor block is given by
\begin{subequations}
\begin{eqnarray}
    \check{\mu}\hat{\tilde{\gamma}}^{TF}_{AB}&=&-2\alpha\hat{\tilde{A}}^{TF}_{AB}~,\\
    \check{\mu}\hat{\tilde{A}}^{TF}_{AB}&=&-\frac{\alpha}{2}\hat{\tilde{A}}^{TF}_{AB}~,
\end{eqnarray}
\end{subequations}
and the vector block reads
\begin{subequations}
\begin{eqnarray}
    \check{\mu}\hat{\tilde{\gamma}}_{\perp A}&=&-2\alpha\hat{\tilde{A}}_{\perp A} +\chi \hat{\beta}_A~,  \\
    \check{\mu}\hat{\tilde{A}}_{\perp A}&=&\frac{\chi \alpha}{2}\big(\chi \hat{\tilde{\Gamma}}_A-\chi ^{-1} \hat{\tilde{\gamma}}_{\perp A}\big)~, \\
    \check{\mu}\hat{\tilde{\Gamma}}_A &=& \chi^{-1} \hat{\beta}_A - \frac{2 \alpha b(x) }{\chi^{2}(1+b(x))}\hat{\tilde{A}}_{\perp A}~, \\
    \check{\mu}\hat{\beta}_A &=& \frac{d}{2(d-1)} \frac{\hat{\tilde{\Gamma}}_A}{1+a(x)}~.
\end{eqnarray}
\end{subequations}
The eigenvalues of the traceless tensor block are $\xi=\pm \alpha$, while those of the vector block are given by
\begin{align}
    \xi &= \left\{\pm \sqrt{\frac{d}{2(1+a(x))(d-1) \chi}}, \quad \pm \frac{\alpha}{\sqrt{1+b(x)}}\right\}.
\end{align}
Note that these eigenvalues are identical to those obtained from the mCCZ4 formalism~\cite{AresteSalo:2023mmd}, and they are degenerate when $\alpha^2 \chi = \frac{d}{2(d-1)}\frac{1+b(x)}{1+a(x)}$.  
Since $\alpha^2 \chi$ is bounded above by 1, we can avoid the degeneracy by choosing 
$\frac{d}{2(d-1)}\frac{1+b(x)}{1+a(x)}>1$, which gives $b(x)> \frac{d-2}{d}+\frac{2(d-1)a(x)}{d}$.  
Nevertheless, for practical applications, we expect that this constraint can be relaxed to  $b(x) \neq \frac{d-2}{d}+\frac{2(d-1)a(x)}{d}$ so that it will not be degenerate at spatial infinity, as in the case for mCCZ4.

Finally, the scalar block is
\begin{subequations}
\begin{eqnarray}
    \check{\mu}\hat{\tilde{\gamma}}_{\perp \perp}&=&-2\alpha\hat{\tilde{A}}_{\perp \perp} +2\chi \frac{d-1}{d}\hat{\beta}^\perp \,, \\
    \check{\mu}\hat{\chi} &=& \frac{2}{d} \chi \big( \alpha \hat{K} - \hat{\beta}^\perp \big)\,, \\
    \check{\mu}\hat{K} &=& - \hat{\alpha} +\frac{\alpha b(x) (d-2)}{2(d-1)(1+b(x))} \big( (d-1) \chi ^{-1} \hat{\chi}\nonumber\\&&\hspace{4cm}+\chi^{-1} \hat{\tilde{\gamma}}_{\perp \perp}\big)\,, \\
    \check{\mu}\hat{\tilde{A}}_{\perp \perp}&=& -\frac{\alpha}{2}\hat{\tilde{\gamma}}_{\perp \perp}+\frac{\alpha (d-1)(d-2)}{2d} \hat{\chi}\nonumber\\&&+ \chi^2 \alpha \frac{d-1}{d} \hat{\tilde{\Gamma}}^\perp -\chi \frac{d-1}{d} \hat{\alpha}\,, \\
    \check{\mu}\hat{\tilde{\Gamma}}^\perp &=& \frac{2(d-1)}{d}\chi^{-1} \hat{\beta}^\perp - 2\alpha \chi^{-2} \hat{\tilde{A}}_{\perp \perp}\nonumber \\&&+\frac{2 \alpha }{\chi^{2}(1+b(x))} \big(\hat{\tilde{A}}_{\perp \perp}-\frac{d-1}{d} \chi \hat{K} \big)\,, \\
    \check{\mu}\hat{\alpha} &=&  -\frac{2 \alpha}{1+a(x)} \hat{K}\,, \\
    \check{\mu}\hat{\beta}^\perp &=& \frac{d}{2(d-1)} \frac{\hat{\tilde{\Gamma}}^\perp}{1+a(x)}-\frac{a(x)}{1+a(x)} \alpha \hat{\alpha}\,,
\end{eqnarray}
\end{subequations}
\\
with eigenvalues  
\begin{align}
    \xi = &\left\{0,~\pm \frac{1}{\sqrt{\chi (1+a(x))}}, \right. \nonumber \\
    &\left. \pm \sqrt{\frac{2 \alpha}{1+a(x)}},~\pm \frac{\alpha }{\sqrt{(1+b(x))}}\right\}~.
\end{align}
The eigenvalues obtained are identical to those found in mCCZ4,\footnote{There is a difference in multiplicity in the last pair of eigenvalues between the mBSSN and mCCZ4 formalisms.} with the addition of a vanishing eigenvalue $\xi=0$. It is well known that this non-propagating mode exists in the standard BSSN formulation for GR~\cite{Beyer:2004sv, Brown:2008sb}, and could lead to greater constraint violations in numerical simulations~\cite{Bernuzzi:2009ex}.
The eigenvalues from the scalar block are degenerate when 
$\alpha = \frac{1}{2 \chi}$, $\alpha ^2 = \frac{1+b(x)}{\chi (1+a(x))}$ and $\alpha = \frac{2(1+b(x))}{1+a(x)}$.  
Similar to mCCZ4 and other well-known formalisms for GR (e.g. CCZ4 and BSSN), these degeneracies do not pose a problem for practical purposes as long as the modified gauge satisfies $a(x) \neq b(x)$, which avoids degeneracy at spatial infinity.

Analogous to \cite{Kovacs:2020pns,Kovacs:2020ywu,AresteSalo:2023mmd}, we may arrange our eigenvalues into three groups:\footnote{This arrangement of eigenvalues can also be seen by following the analysis done in~\cite{Brown:2008sb} for BSSN.}
\begin{itemize}
    \item Physical eigenvalues: $\xi=\pm \alpha$ with multiplicity $d-1$ for each sign.
    \item Gauge-condition violating eigenvalues: $\xi = \pm \sqrt{\frac{2\alpha}{1+a(x)}},\pm \frac{1}{\sqrt{\chi(1+a(x))}}, \pm \sqrt{\frac{d}{2(1+a(x))(d-1)\chi}}$ with the multiplicity of the last pair being $d-1$.
    \item Pure-gauge eigenvalues: $\xi=0, \pm\frac{\alpha}{\sqrt{1+b(x)}}$ with the multiplicity of the last pair being $d-1$.
\end{itemize}

Their corresponding eigenvectors have been explicitly written in Appendix~\ref{app:eigenvectors}. Clearly the eigenvalues are real (recall that in all cases $a(x)>-1$ and $b(x)>0$), they smoothly depend on $n_i$ and so do the corresponding eigenvectors. Therefore, we conclude that the mBSSN formalism is strongly hyperbolic in GR, and hence is well-posed.  

We will now extend this result to discuss the well-posedness of the mBSSN formalism in beyond-GR theories. In particular, we will limit ourselves to the theories that are known to be well-posed in mCCZ4, namely Einstein-Gauss-Bonnet gravity and $4\partial$ST theories. These theories contain couplings to higher-order terms or additional degrees of freedom (e.g., scalar fields), have (completely non-linear) second-order equations of motion, and are continuously connected to GR (i.e., they reduce to GR in the zero coupling limit). Note that when considering additional fields (such as scalar fields in $4\partial$ST theories), we should, in principle, analyse the full Einstein-scalar system instead of just Einstein's equations, as the theory reduces to GR with the addition of a minimally coupled scalar in the zero coupling limit. However, since the principal symbol of the scalar field equation forms its own block within the scalar block during the hyperbolicity analysis and is untouched by the modified gauge (see e.g.,~\cite{AresteSalo:2023mmd}), it will only contribute two additional physical eigenvalues with $\xi = \pm \alpha$. We therefore do not include their contributions explicitly in the calculations above.
\subsection{Well-posedness beyond GR}
\label{sec:bGR_wellposedness}
We have demonstrated that the mBSSN and mCCZ4 formalisms share the same set of eigenvalues in GR (with the addition of the $\xi=0$ eigenvalue, which we have justified above). We will now argue that this is sufficient to guarantee well-posedness in modified gravity theories that are known to be well-posed in the mCCZ4 formalism (i.e., Einstein-Gauss-Bonnet gravity and $4\partial$ST theories) in the weakly coupled regime.

Note that one can encode the contribution from the modified gravity terms in the equation of motion as an effective $T^{\mu\nu}$ tensor. The effective $T^{\mu\nu}$ decomposes as usual into $\rho$, $S_i$, and $S_{ij}$ [see Eq.\,\eqref{eqsTmunu}], but now these terms can contain second-order derivatives of the metric and the additional degrees of freedom and, thus, change the principal part of the equations. Hence, the well-posedness analysis in GR is no longer valid for certain classes of modified gravity theories.

Nevertheless, the key insight behind~\cite{Kovacs:2020pns, Kovacs:2020ywu} is that by separating the propagation speeds of the ``physical'', ``pure gauge'' and ``gauge-condition-violating'' modes in GR, one prevents the modes from mixing and causing degeneracies, which is the reason for the loss of hyperbolicity in certain classes of modified theories~\cite{Papallo:2017qvl, Papallo:2017ddx} when analysed using standard GR tools.\footnote{It is also possible, most likely away from the weak coupling regime, that the loss of hyperbolicity is due to the physical degrees of freedom. See, for example~\cite{R:2022hlf,Doneva:2023oww,Thaalba:2024crk}.} In the modified gauge, as one deforms the theory to introduce corrections to GR, one can group the new modes of the modified theory by the corresponding GR modes they reduce to in the zero coupling limit. In this way, we can divide the modes of the modified theory into the ``physical'', ``pure gauge'' and ``gauge-condition-violating'' sectors of GR, which are independent of each other.  This allows one to analyse the hyperbolicity of the system by studying each group independently. Since mBSSN can separate the three sectors in GR, we will now argue for each sector how the results from mCCZ4 are expected to remain valid here.

We first examine the sector corresponding to the physical modes. By definition, the physical modes are gauge-independent. The introduction of the modified gauge also removes the possibility of mixing with gauge modes.
Therefore, the analysis carried out in mCCZ4 regarding the physical sector remains equally valid in mBSSN. We then examine the sectors corresponding to the ``pure gauge'' and ``gauge-condition-violating'' modes. As discussed in~\cite{AresteSalo:2023mmd} for the case of mCCZ4, the non-trivial corrections due to modified gravity only impact the physical modes. Therefore, if the ``pure gauge'' and ``gauge-condition-violating'' modes are well-behaved in GR, they will remain so in the weakly coupled modified theories we consider. Given the similar structures of mBSSN and mCCZ4 as reflected by their eigenvalues and the identical degeneracy conditions for the eigenvectors, this property remains equally valid for the mBSSN, implying that the ``pure gauge'' and ``gauge-condition-violating'' modes of mBSSN remain well-behaved in the modified theories we consider.
The above arguments suggest that the analysis performed for mCCZ4 can be suitably adapted to our formalism, provided that the three types of modes are separated at the level of GR, which is satisfied by mBSSN.  This indicates that mBSSN is well-posed for beyond-GR theories which are known to be well-posed in mCCZ4. We will test this numerically below.

\section{Modified BBSN formalism: Numerical studies}\label{sec:numerics}
In this section, we present the numerical setup and the relevant physical quantities extracted and analysed from the simulations. In particular, we compare the performance of the mBSSN and the mCCZ4 formalisms by studying two different systems: (i) an isolated spinning BH and (ii) a head-on BH merger in Einstein-scalar-Gauss-Bonnet (EsGB) theory. More precisely, we will consider the following action for the EsGB theory:
\begin{equation}
    S=\frac{1}{16\pi G}\int d^4x \sqrt{-g} \left(R-\frac{1}{2}\partial_\mu\phi\,\partial^\mu \phi+\lambda \, f(\phi)\, R_{\rm GB}\right)~,
\end{equation}
where $R_{\rm GB}=R^2-4R^{\mu \nu}R_{\mu \nu} + R^{\mu \nu \rho \sigma}R_{\mu \nu \rho \sigma}$ is the Gauss-Bonnet curvature, $\lambda$ is the dimensionful coupling constant and $f(\phi)$ is an arbitrary coupling function. We will take this function to be linear  
$f(\phi)=\phi$, which corresponds to a shift-symmetric theory \cite{Sotiriou:2013qea}.
The linear coupling of the scalar to the Gauss-Bonnet term circumvents the no-hair theorem of~\cite{Hui:2012qt}, and leads to hairy BHs~\cite{Sotiriou:2013qea, Sotiriou:2014pfa}.  

\subsection{Numerical setup and initial data}\label{subsec:setup}
We implement the new mBSSN equations into the public code \texttt{GRFolres}~\cite{AresteSalo:2023hcp}, which is an extension of \texttt{GRChombo}~\cite{Clough:2015sqa, Andrade:2021rbd} for evolving modified gravity theories using the mCCZ4 formalism. The grid settings used for both systems are summarised in Table~\ref{table:numerical_settings}.
\begin{table}[h!]
    \centering
    \begin{tabular}{|l|c|c|c|}
        \hline
        \textbf{System} & $N$ & $L/M$ & AMR levels \\
        \hline
        Spinning BH     & 128          & 512          & 8                   \\
        \hline
        Head-on merger  & 96           & 512          & 10                   \\
        \hline
\end{tabular}
\caption{\emph{Numerical setup.} Numerical settings for the isolated spinning BH and head-on merger simulations, respectively. The number of grid points at the coarsest level is denoted as $N$, the size of the box is denoted as $L$, and the final column denotes the number of adaptive mesh refinement (AMR) levels (including the coarsest level).}
\label{table:numerical_settings}
\end{table}
For both systems, we will use the same choice of modified gauge with $(a,b)=(0.2,0.4)$. The coupling constant is taken to be $\lambda=0.1 m_1^2$.\footnote{Here, $m_1$ is the scale defined by the mass of the smallest BH in the system, which is not always equal to the arbitrary code unit $M$ (which we set to be the total ADM mass of the spacetime).  For example, in an equal mass binary BH system, we will have $m_1=0.5M$.} The scalar field and its conjugate momentum will be taken to be initially vanishing, such that the constraint equations reduce to those of GR, allowing us to make use of GR initial data. Using this initial data will lead to a short transient behaviour at the start of the simulation as the scalar field relaxes to a configuration containing hairy BHs; however, this transient timescale is much shorter than the orbital timescale of the BHs. Note that this approach has been widely used in studying binary systems in EsGB gravity (see e.g.~\cite{Corman:2022xqg,AresteSalo:2025sxc,Corman:2025wun}). We will therefore follow the same approach here. We note that progress has been made in the community towards constructing accurate initial data for modified theories of gravity~\cite{Brady:2023dgu,Nee:2024bur}.

We will perform simulations using both the mBSSN and mCCZ4 formalisms to benchmark the results. We will extract and compare several diagnostics from the simulations. First, we extract the gravitational and scalar waves using the standard Newman-Penrose formalism~\cite{Newman:1961qr,Bishop:2016lgv}. In this approach, one defines a suitable null tetrad that provides the vectors for constructing different projections of the Weyl tensor, yielding a collection of complex scalars. One of these scalars,  $\Psi_4$, encodes the outgoing radiative degrees of freedom. This will be the scalar that we extract from our simulations to study the gravitational waves. In particular, we will extract the $\Psi_4$ scalar projected into multipoles on a sphere of fixed radius, i.e., 
\begin{equation}
    r_{\text{ex}}\Psi_{4,lm}=\int d\Omega \,r_{\text{ex}}\Psi_{4}|_{r=r_{\text{ex}}} \prescript{}{-2}{Y_{lm}^*}~,
\end{equation}
where $\prescript{}{-2}{Y_{lm}}$ are the spin-weighted spherical harmonics. A similar procedure is performed for the extraction of scalar waves, where the scalar is decomposed into multipoles using spherical harmonics:
\begin{equation}
    r_{\text{ex}}\phi_{lm}=\int d\Omega \,r_{\text{ex}}\phi|_{r=r_{\text{ex}}} Y_{lm}^*~.
\end{equation}
In the following, we will use $(l,m)$ to denote any particular projections of the $\Psi_4$ and $\phi$. For the isolated spinning BH, we first extract the scalar waves generated throughout the simulation. From the geometry of our setup, the dominant scalar mode is the $(0,0)$ mode, and we will extract this to compare the formalisms. We will also extract the apparent horizon (AH) mass $M_{\text{AH}}$ and spin $a_{\text{AH}}$ of the BH as measured assuming GR via Christodoulou formula \cite{Christodoulou:1970wf}.\footnote{We note that both the mass and angular momentum computed at the horizon do not coincide with its ADM counterparts when the coupling constant is non-zero given that the scalar sector carries away some part of the mass (see discussion for a non-spinning BH in the Supplemental Material of \cite{Corman:2025wun} and references therein).} 

Finally, we will extract the Hamiltonian constraint violation from our simulation (see Eq.~\eqref{Ham_eq}). Note that the Hamiltonian constraint is not expected to be satisfied within the AH. To remove such contributions, we use the conformal factor $\chi$ to identify the approximate location of the horizon~\cite{Radia:2021smk}. The extracted Hamiltonian is then rescaled by $\mathcal{H} \to \frac{\mathcal{H}}{1+e^{-1000(\chi -\chi_0)}}$, where $\chi_0=0.2\sqrt{1-(a/M)^2}$ is the conformal factor at a location slightly within the horizon and $a$ is the spin of the BH, such that the Hamiltonian is extracted as normal outside but is heavily suppressed within the horizon.

For the head-on merger, we will extract 
the $(2,2)$ mode of the $\Psi_4$ scalar and the $(0,0)$ mode of the scalar wave.  To exploit the problem's symmetries, we apply reflective boundary conditions on the appropriate boundaries. In such dynamical systems, one can also observe a dephasing of the gravitational waves when comparing simulations of modified gravity with GR. However, it is not clear how this is affected by the initial transient as pointed out in the Ref.~\cite{Corman:2025wun}.
To be precise, we will perform four sets of simulations for the head-on merger scenario: (i) EsGB gravity with mBSSN formalism, (ii) EsGB gravity with mCCZ4 formalism, (iii) GR with mBSSN formalism (with the modified gauge activated as before), and (iv) GR with mCCZ4 formalism. The results for both scenarios are summarised in the following sections.

\subsection{Isolated spinning BH}
\begin{figure}
    \centering
    \includegraphics[width=1\linewidth]{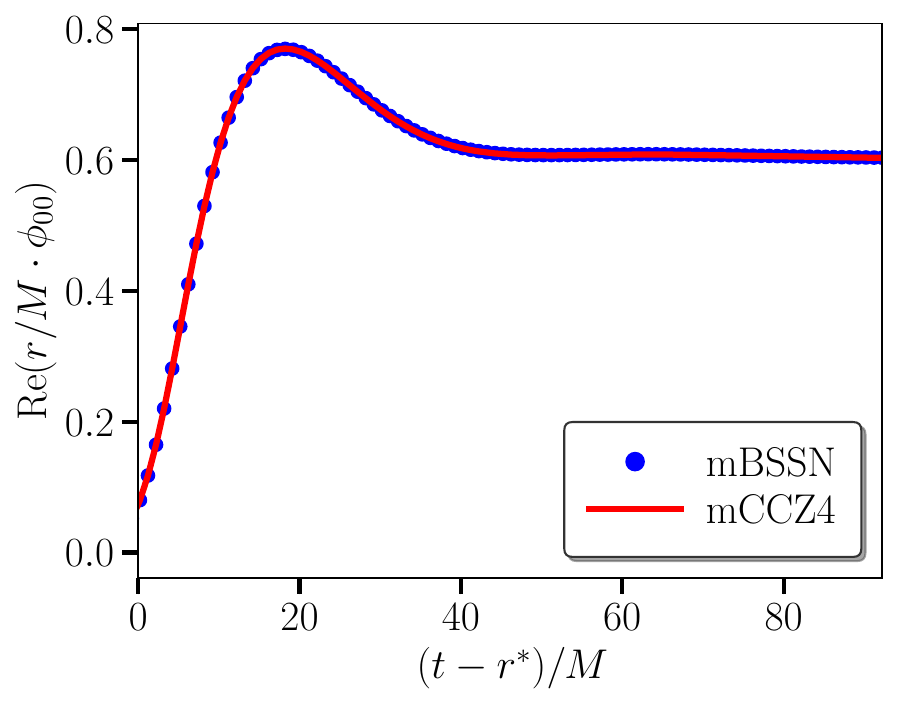}
    \caption{\emph{Scalar waves of an isolated spinning BH.} The real part of the $(0,0)$ mode of the scalar field $\phi$ extracted at $r=100M$.  The scalar waves from both mBSSN and mCCZ4 are compared, showing a strong agreement between the formalisms.}
    \label{fig:SW00}
\end{figure}
Here we consider an isolated spinning BH of spin $a=0.7 M$, which coincides with the typical value of the spin of the remnant in a quasi-circular initially non-spinning binary BH merger. In Fig.~\ref{fig:SW00}, we present the $(0,0)$ mode of the scalar wave extracted at $r=100M$ as a function of retarded time $t^* \equiv t-r^*$, where $r^*=r+2M\ln\left(\frac{r}{2M}-1\right)$ is the tortoise coordinate evaluated at the radius of extraction. Both formalisms appear to produce consistent results and predict the same final scalar charge for the BH.
\begin{figure}
    \centering
    \includegraphics[width=1\linewidth]{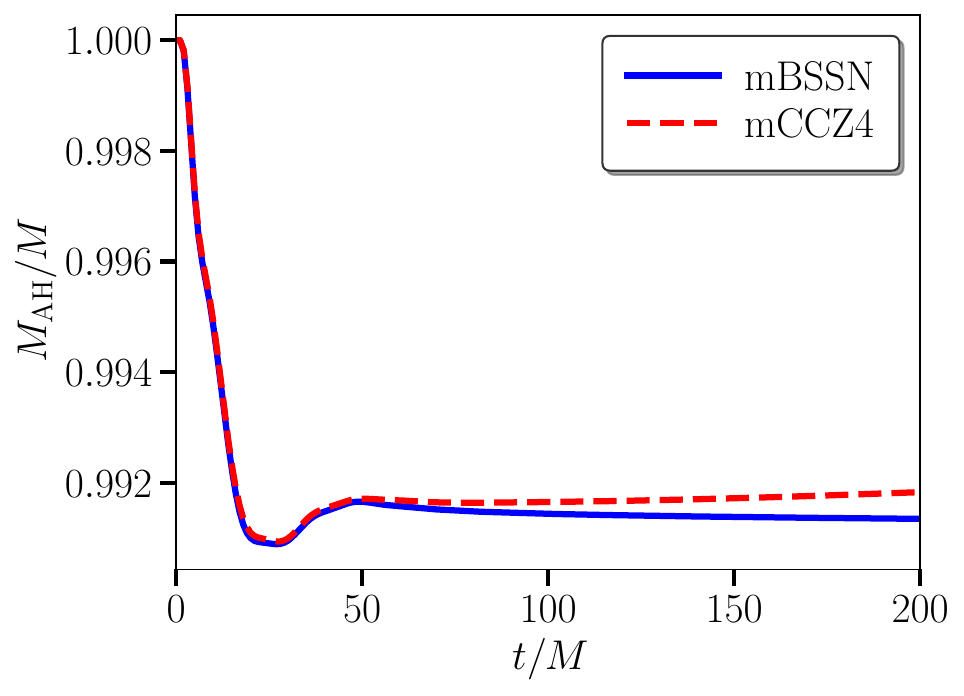}
    \caption{\emph{Apparent horizon mass of an isolated spinning BH.} The AH mass of the BH as a function of time. We compare the mass of the BH as measured at the apparent horizon for both formalisms. We find that the mass experiences a slight increase at late times for the mCCZ4 formalism, while the mBSSN formalism exhibits a slight decrease.}
    \label{fig:AHmass}
\end{figure}
\begin{figure}
    \centering
    \includegraphics[width=1\linewidth]{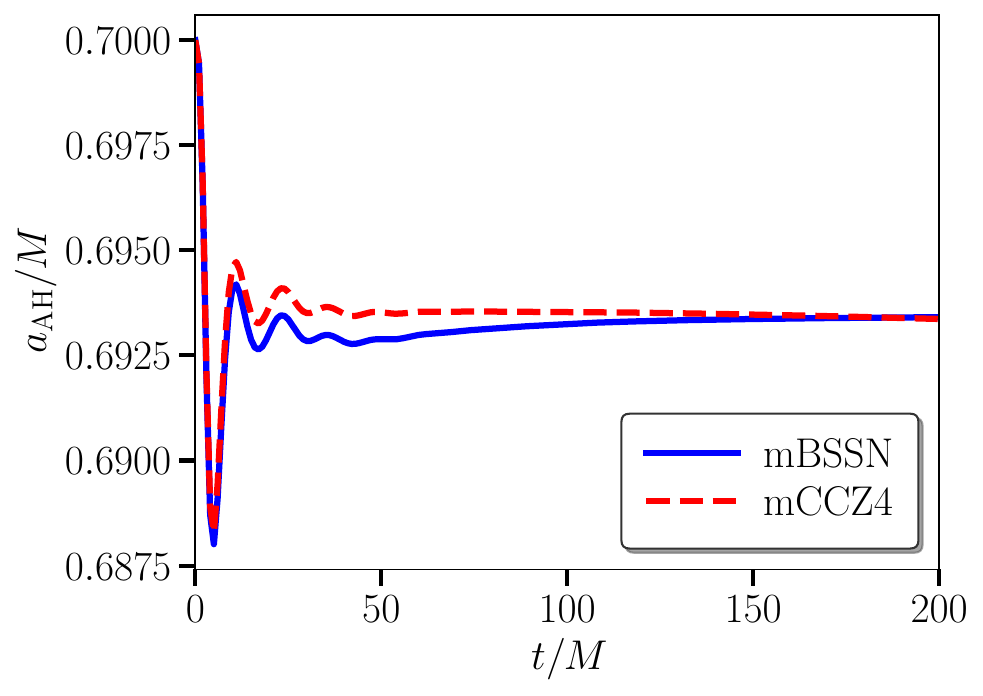}
    \caption{\emph{Apparent horizon spin of an isolated spinning BH.} The AH spin of the BH as a function of time. We compare the spin of the BH as measured at the apparent horizon for both formalisms. We find that mBSSN agrees better with mCCZ4 at late times, though one can still observe a slight increase in mBSSN and decrease in mCCZ4.}
    \label{fig:AHspin}
\end{figure}

The AH mass and spin, as shown in Figs.~\ref{fig:AHmass}, and~\ref{fig:AHspin}, are also broadly consistent.  Note that the decrease in AH mass at the beginning of the simulation is consistent with the expectation that part of the BH mass gets transferred to the scalar sector as the scalar hair develops.  A similar argument holds for the AH spin, where the BH loses angular momentum to the scalar field.  We note that analogous results have been obtained in \cite{East:2020hgw, AresteSalo:2023mmd}. At late times, the AH mass obtained from the mCCZ4 formalism experiences a slight increase, whereas the mBSSN formalism experiences a slight decrease. For the AH spin, mBSSN and mCCZ4 agree better at late times, though one can still observe a slight decrease in mCCZ4 and increase in mBSSN.  The behaviour of the mCCZ4 simulations can in principle be improved by choosing a different set of damping parameters.  On the other hand, it is not known whether varying the modified gauge would lead to better or worse performance in these diagnostics for both formalisms; this will be studied in more detail in future work. 
\begin{figure}
    \centering
    \includegraphics[width=1\linewidth]{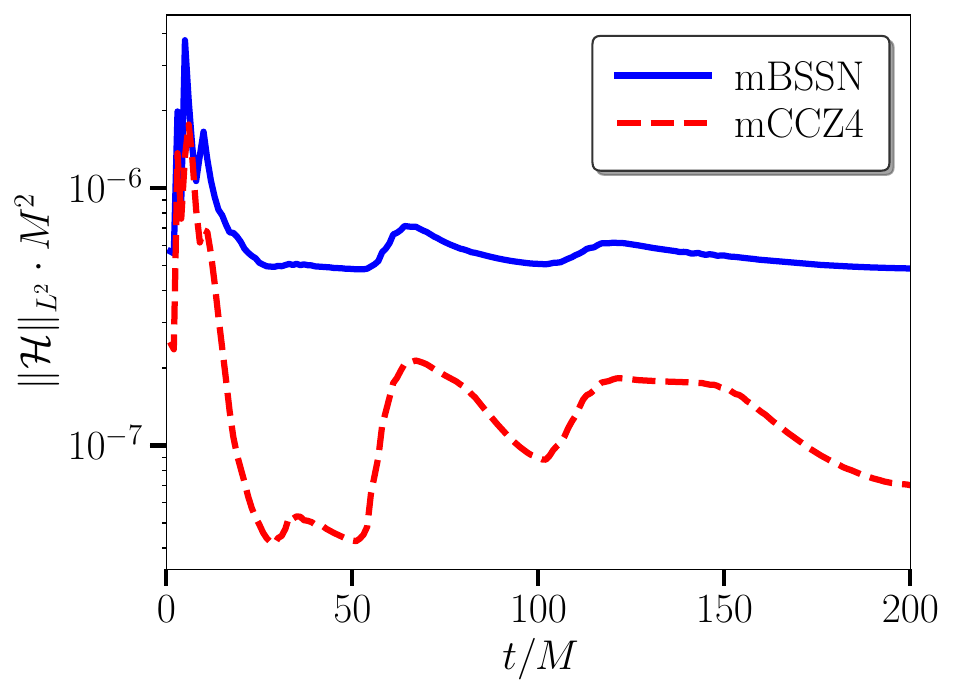}
    \caption{\emph{Hamiltonian constraint violation of an isolated spinning BH.} 
    $L^{2}$ norm of the Hamiltonian, $\mathcal{H}$, across the numerical grid as a function of time. The mBSSN formulation shows slightly larger Hamiltonian constraint violation than mCCZ4, although the latter remains approximately constant at late times. This trend is consistent with previous comparisons between the standard BSSN and CCZ4 formulations~\cite{Sanchis-Gual:2014nha}.}
    \label{fig:Ham}
\end{figure}

Finally, we compare the Hamiltonian constraint violations of the mBSSN and mCCZ4 formalisms. As shown in Fig.~\ref{fig:Ham}, the mBSSN has a larger Hamiltonian constraint violation when compared to mCCZ4, likely due to the absence of constraint damping terms in the formalism. Despite this, the constraint violations stay relatively constant over time and do not exhibit significant growth. These results are similar to those obtained in studies comparing the performance of the standard BSSN and CCZ4 in GR~\cite{Bernuzzi:2009ex,Alic:2011gg,Sanchis-Gual:2014nha}.
\subsection{Head-on mergers}
\begin{figure*}
    \centering
    \includegraphics[width=\textwidth]{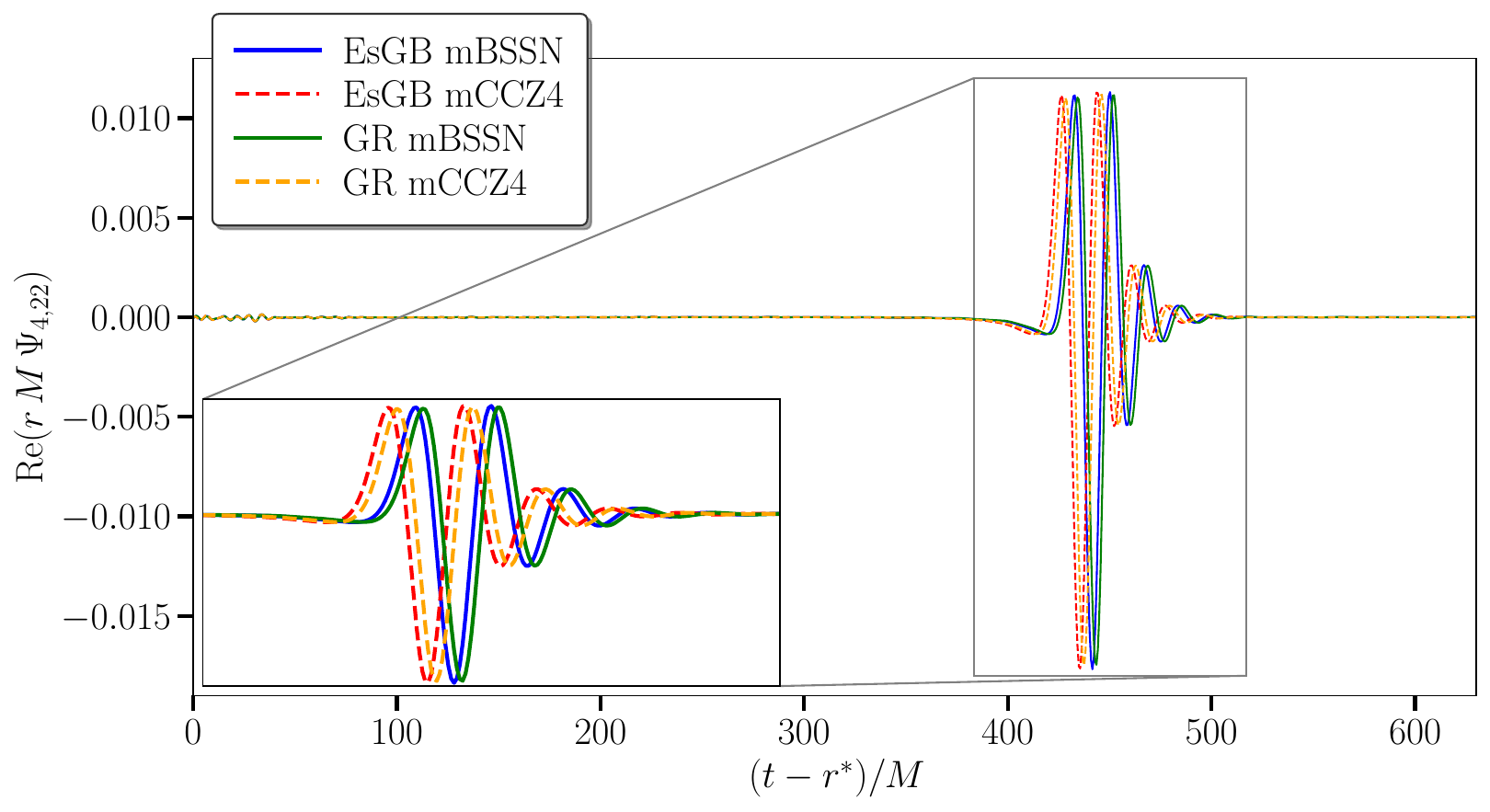}
    \caption{\emph{Gravitational waves of a head-on merger.} The real part of the main $(2,2)$ mode of the Newman-Penrose scalar $\Psi_{4},$ as a function of time, extracted at $r=100M$, for GR and EsGB simulations, using the mCCZ4 and mBSSN formalisms, respectively. We find here a very good agreement between both formalisms and especially in the observed dephasing between GR and non-GR.}
    \label{fig:headon_weyl22}
\end{figure*}
\begin{figure}
    \centering
    \includegraphics[width=1\linewidth]{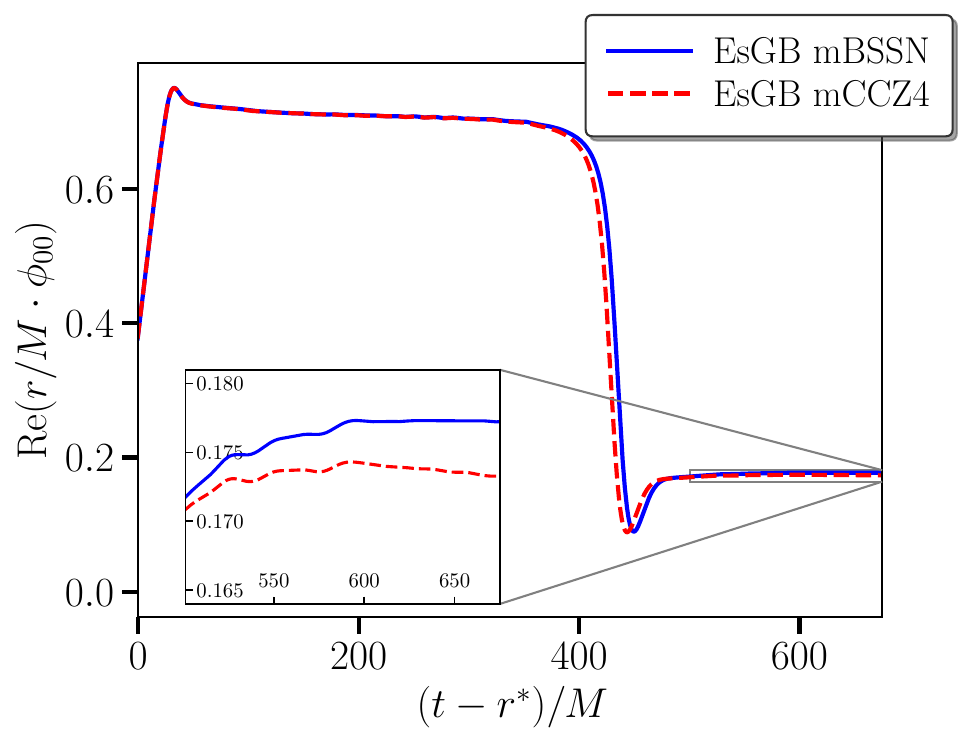}
    \caption{\emph{Scalar waves of a head-on merger.} The scalar field $(0,0)$ mode, as a function of time,  extracted at $r=100M$. The scalar field grows and eventually settles to a smaller value as the BHs merge because of the larger mass of the remnant compared to that of the pre-merger BHs. The waveform of the $(0,0)$ mode is largely consistent between both formalisms, with a minor difference in merging time (as reflected in previous figures) and the final value of the scalar field.  The inset shows the small deviation in the final scalar charge between the two formalisms.
    }
    \label{fig:headon_scalar00}
\end{figure}
Here, we initially place two equal-mass, non-spinning BHs at a separation of $50M$, and evolve the system past the merger and ringdown. In Fig.~\ref{fig:headon_weyl22}, we show the gravitational waves as reflected by the real part of the $(2,2)$ mode of the $\Psi_4$ scalar. We find that mCCZ4 simulations merge at approximately $10M$ earlier than mBSSN simulations, likely due to the presence of additional dissipation inherent in the formalism. The difference in gauge settlement at the start of the simulation can also contribute to this difference. 
Within each formalism, we find consistently that the EsGB simulations merge slightly earlier than the GR simulations, with a very similar 
amount of dephasing for both formulations, although it is not clear how this depends on the initial transient, as we pointed out earlier.
The amplitude of the waves and the dephasing between GR and EsGB are also consistent for both formalisms.

For the scalar sector, we again extract the $(0,0)$ mode of the scalar wave (emission of scalar waves through higher modes is not significant under our current setup). The result is shown in Fig.~\ref{fig:headon_scalar00}. We find that both formalisms agree on the total scalar charge of the spacetime initially and immediately after the merger (the ratio between the amplitudes pre and post-merger corresponds approximately to a factor of $M^2/(m_1m_2)=4$ as analytically expected), with minor differences arising at late times. After around $t^* \sim 600M$, the scalar charge predicted by mCCZ4 experiences a slight decrease, while mBSSN stays relatively constant.
Similar to the above, whether this can be improved by choosing a different modified gauge will be explored in future work.
\section{Discussion}
In GR, several $d+1$ formalisms, such as BSSN, Z4, Z3, and CCZ4, have been developed to enable numerical studies of physically interesting scenarios.  It is also known that these formalisms are mathematically related to each other. 
Given that the mCCZ4 formalism has proven successful for certain classes of beyond-GR theories, it is natural to consider modified versions of the other formalism and explore whether there is a similar link between them beyond GR.
We have shown that this is indeed the case and demonstrated how one can start from the mCCZ4 formalism and derive a modified BSSN formalism.

We explicitly showed that mBSSN is well-posed in GR and argued that it remains well-posed for weakly coupled modified theories known to be well-posed in mCCZ4, such as Einstein-Gauss-Bonnet gravity and $4\partial$ST theories.  We then tested the robustness of the new formalism by performing numerical experiments involving an isolated spinning BH and a head-on merger of two BHs in EsGB gravity.  The numerical results were benchmarked using the mCCZ4 formalism, where we found strong agreement between the two. 

In addition to the above results, we observe that the relative performance of the mBSSN and mCCZ4 formulations in EsGB gravity reproduces what has been previously established in pure GR~\cite{Alic:2011gg,Sanchis-Gual:2014nha}. Earlier comparative studies of BSSN and CCZ4 found that CCZ4 typically leads to better constraint damping and improved stability. At the same time, BSSN remains competitive in accuracy and robustness (see Ref.~\cite{Bernuzzi:2009ex} for a similar comparison and result comparing Z4c against BSSN). 

In our current study, we restricted ourselves to a single choice of (constant) parameters for the modified gauge.  It is not known whether a different choice of gauge parameters would improve numerical performance.  For example, a better choice of gauge parameters might lead to improved numerical stability, as reflected in the late-time behaviour of the diagnostic quantities we extract from the simulations.  One may also, in principle, generalise the gauge condition by promoting the constant parameters to position-dependent functions.  Another possible extension of this work is to investigate whether each formalism is best suited for studying different scenarios.  This will be left for future work.

Finally, the results presented here also open up exciting opportunities for the broader numerical relativity community. Since both the mBSSN and mCCZ4 formalisms can be written in a way closely paralleling their GR counterparts, they offer a natural pathway for extension to existing numerical relativity infrastructures. In particular, NR codes such as the Einstein Toolkit~\cite{Loffler:2011ay,canuda}, SACRA~\cite{Yamamoto:2008js}, BAM~\cite{Bruegmann:2003aw}, SPHINCS\_BSSN~\cite{Rosswog:2020kwm} and Dendro~\cite{Fernando:2022php} could adopt these modified formalisms with relatively modest adaptations. This would greatly expand the range of modified theories of gravity that these codes can simulate, enabling systematic studies of beyond-GR effects in binary mergers and gravitational-wave signatures.

\section*{Acknowledgments}
We acknowledge helpful conversations with Fernando \'Abalos and Katy Clough. We thank the GRTL collaboration (\url{www.grtlcollaboration.org}) for their support and code development work. HLHS
is supported by the Croucher Scholarship. FT is supported by the INFN Post-doctoral research agreement No. 27076. MB acknowledges partial support from the STFC Consolidated Grant nos. ST/Z000424/1 and UKRI2492. TPS acknowledges partial support from the STFC Consolidated Grant nos. ST/V005596/1,  ST/X000672/1, and UKRI2492. This work used the DiRAC Memory Intensive service Cosma8 at Durham University, managed by the Institute for Computational Cosmology on behalf of the STFC DiRAC HPC Facility (\url{www.dirac.ac.uk}). The DiRAC service at Durham was funded by BEIS, UKRI and STFC capital funding, Durham University and STFC operations grants. DiRAC is part of the UKRI Digital Research Infrastructure. We also used the resources and services provided by the VSC (Flemish Supercomputer Center), funded by the Research Foundation - Flanders (FWO) and the Flemish Government.

\appendix
\section{Eigenvectors of the principal part of Einstein's equations}
\label{app:eigenvectors}
In this appendix, we present the eigenvectors of the principal part of Einstein's equations~\eqref{eqsbssn}. The eigenvectors are grouped according to the nature of their eigenvalues, i.e., whether they belong to the ``physical'', ``pure gauge'' or the ``gauge-condition violating'' modes.
\begin{widetext}

\vspace{-0.4cm}

\begin{table}[h!]
\centering
\begin{tabular}{|c|c|c|c|}
\hline
$\hat{\gamma}_{ij}$ & $\hat{A}_{ij}$  \\
\hline
$0$ & $0$ \\
$\mp 2 e_A^i e_B^j$ & $e_A^i e_B^j$  \\
$\pm 2(e_A^i e_A^j - e_B^i e_B^j)$ & $-e_A^i e_A^j + e_B^i e_B^j$ $ \quad \forall A \neq B$ \\
\hline
\end{tabular}
\caption{Physical eigenvectors of the GR principal part.}
\end{table}

\vspace{-0.5cm}
    
\begin{table}[h!]
\centering
\fontsize{7pt}{7pt}\selectfont
\begin{tabular}{|c|c|c|c|}
\hline
$\hat{\gamma}_{ij}$ & $\hat{\chi}$ & $\hat{\alpha}$& $\hat{\beta}^i$ \\ 
$\hat{\tilde{A}}_{ij}$ & $\hat{K}$ & $\hat{\Gamma}^i$ & \\ 

\hline
$\mp \sqrt{\frac{2(1+a(x))(d-1)\chi^3}{d}}e^i_A n^j$ & $0$ & $0$ & $e^A_i$\\
$0$ & $0$ & $\mp \sqrt{\frac{2(1+a(x))(d-1)}{d\chi}}e^i_A$ &\\
\hline
$ \mp \frac{2 \sqrt{a(x)+1} (d-1) \chi ^{3/2}}{d} n^in^j$ & $\pm \frac{2 \sqrt{a(x)+1} \chi ^{3/2}}{d}$ & $0$ & $n_i$ \\
$0$ & $0$ & $\mp \frac{2 \sqrt{a(x)+1} (d-1)}{d \sqrt{\chi }}n_i$ &\\
\hline
$ \pm \frac{2 \sqrt{2 \alpha (a(x)+1)} (d-1) \chi ^2}{d (a(x) (2 \alpha  \chi -1)-1)}n^in^j$ & $\pm \frac{2 \sqrt{2 \alpha (a(x)+1)} \chi ^2}{-2 a(x) \alpha  d \chi +d(1+a(x)}$  & $\frac{\sqrt{2(a(x)+1)} (2 \alpha  \chi -1)}{\sqrt{\alpha } (a(x) (2 \alpha  \chi -1)-1)}$& $n_i$\\
$\frac{(a(x)+1) (d-1) \chi  (2 \alpha  \chi -1)}{\alpha  d (a(x) (2 \alpha  \chi -1)-1)}n^in^j$ &
$\frac{(a(x)+1) (2 \alpha  \chi -1)}{\alpha  (a(x) (2 \alpha  \chi -1)-1)}$& $\pm \frac{2 \sqrt{2 \alpha (a(x)+1)} (d-1)}{d (a(x) (2 \alpha  \chi -1)-1)}n_i$ &\\
\hline
\end{tabular} \\
\caption{``Gauge-condition violating'' eigenvectors.}
\end{table}

\vspace{-0.5cm}

\begin{table}[h!]
\centering
\begin{tabular}{|c|c|c|c|}
\hline
$\hat{\gamma}_{ij}$ & $\hat{\chi}$ & $\hat{\alpha}$ & $\hat{\beta}^i$ \\ 
$\hat{\tilde{A}}_{ij}$ & $\hat{K}$ & $\hat{\Gamma}^i$ & \\ 
\hline

$\mp \frac{\sqrt{1+b(x)}\chi (-d+2(1+a(x))(d-1)\alpha^2 \chi)}{b(x)d \alpha}e^i_A n^j$ & $0$  & $0$& $e^A_i$\\ 
$\frac{\chi (d(1+b(x))-2(1+a(x))(d-1)\alpha^2 \chi)}{2 b(x)d\alpha}e^i_A n^j$ & $0$ &$\mp \frac{2\alpha(1+a(x))(d-1)}{d \sqrt{1+b(x)}}e_i^A$ & \\
\hline
$\chi  \left(\frac{2 a(x) \alpha  (d-1) \chi }{d}+\frac{1}{\alpha  b(x)}\right)n^in^j$ & $\frac{\chi  (2 b(x) (d-1)+d)}{\alpha  b(x) (d-2) (d-1)}-\frac{2 a(x) \alpha  \chi ^2}{d}$ & $0$ & $n_i$\\ 
$0$ & $0$ & $\frac{2 a(x) \alpha  (d-1)}{d} n_i$ & \\
\hline
$v_{\gamma}n^in^j$& $v_{\chi}$ & $v_{\alpha}$ & $n_i$\\
$v_{A}n^in^j$ &  $v_K$ & $v_{\Gamma}n_i$ & \\
\hline
\end{tabular}
\caption{``Pure-gauge'' eigenvectors.}
\end{table}
\vspace{-0.3cm}
where
{\scriptsize \begin{eqnarray}
   \hspace{-0.85cm}v_{\gamma}&=&\mp \frac{2 \sqrt{b(x)+1} (d-1) \chi  \left(\alpha ^2 \chi  \left(2 (b(x)+1) (a(x) (b(x)-1) d-2 a(x) b(x)-d)-(a(x)+1)^2 \alpha  (b(x)(d-2)-d)\right)-(a(x)+1) \alpha  d+2 (b(x)+1) d\right)}{\alpha  b(x) d \left(2 \alpha  (b(x)+1) (-2 \alpha  a(x) \chi +a(x)+1)+d \left(2 a(x) \alpha ^2 (b(x)+1) \chi -b(x) (a(x) \alpha +\alpha +2)-2\right)\right)}\,,\\
     \hspace{-0.85cm}v_{\chi}&=&\pm \frac{2 \sqrt{b(x)+1} \chi  \left(d (a(x) \alpha +\alpha -2 b(x)-2)-\alpha ^2 (d-2) \chi  \left((a(x)+1)^2 \alpha -2 a(x) (b(x)+1)\right)\right)}{\alpha  d \left(2 \alpha  (b(x)+1) (-2\alpha  a(x) \chi +a(x)+1)+d \left(2 a(x) \alpha ^2 (b(x)+1) \chi -b(x) (a(x) \alpha +\alpha +2)-2\right)\right)},\\
     \hspace{-0.85cm}v_{\alpha}&=&\mp \frac{2 \sqrt{b(x)+1} (d-2) \left(-\left((a(x)+1) \alpha ^2 \chi \right)+b(x)+1\right)}{2 \alpha  (b(x)+1) (-2 \alpha  a(x) \chi +a(x)+1)+d \left(2 a(x) \alpha ^2 (b(x)+1) \chi -b(x) (a(x)\alpha +\alpha +2)-2\right)}\,,\\
     \hspace{-0.85cm}v_A&=&\mp \frac{(d-1) \chi  \left(-\left((a(x)+1) \alpha ^2 \chi \right)+b(x)+1\right) ((a(x)+1) \alpha  (b(x) (d-2)-d)+2 (b(x)+1) d)}{\alpha  b(x) d \left(2 \alpha  (b(x)+1) (-2 \alpha  a(x) \chi   +a(x)+1)+d \left(2 a(x) \alpha ^2 (b(x)+1) \chi -b(x) (a(x) \alpha +\alpha +2)-2\right)\right)}\,,\\
     \hspace{-0.85cm}v_K&=&\frac{(a(x)+1) (d-2) \left((a(x)+1) \alpha ^2 \chi -b(x)-1\right)}{2 \alpha  (b(x)+1) (-2 \alpha  a(x) \chi +a(x)+1)+d \left(2 a(x) \alpha ^2 (b(x)+1) \chi -b(x) (a(x) \alpha +\alpha +2)-2\right)}\,,\\
    \hspace{-0.85cm} v_{\Gamma}&=&\pm \frac{2 \alpha  (d-1) \left(\alpha  (a(x)+1)^2 (b(x) (d-2)-2)-2 (b(x)+1) d (a(x) b(x)-1)+4 a(x) (b(x)+1)^2\right)}{\sqrt{b(x)+1} d \left(2 \alpha  (b(x)+1) (-2 \alpha  a(x) \chi +a(x)+1)+d \left(2  a(x) \alpha ^2 (b(x)+1) \chi -b(x) (a(x) \alpha +\alpha +2)-2\right)\right)}\,.
\end{eqnarray}}
\end{widetext}

\clearpage

\section{Convergence testing}
We perform a convergence test for one of the cases studied numerically using the mBSSN formalism, namely the isolated spinning BH.
To perform the test, we use the average value of the scalar field over the apparent horizon, $\langle\phi\rangle_{\rm AH}$, as a diagnostic to compare across resolutions.

We consider three different resolutions with the number of grid points at the coarsest level being $N_{\rm LR}=128$, $N_{\rm MR}=160$, and $N_{\rm HR}=192$ for the low (LR), medium (MR) and high resolution (HR) runs, respectively. At the same time, the remaining parameters are kept as described in Section~\ref{subsec:setup}.

\begin{figure}
    \centering
    \includegraphics[width=1\linewidth]{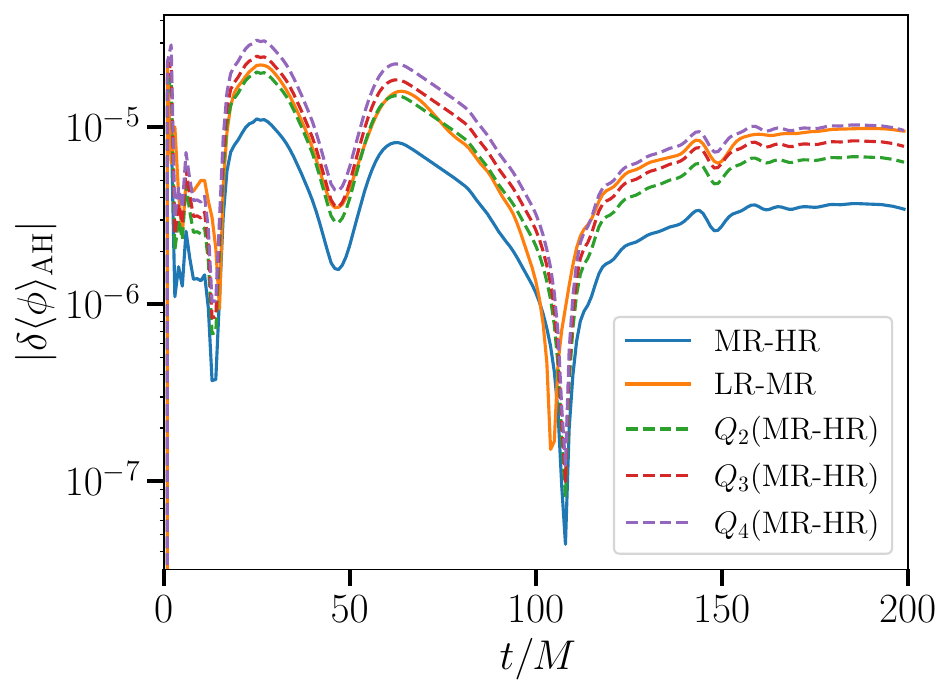}
    \caption{\textit{Convergence assessment.}  We show the difference in the averaged scalar field over the apparent horizon between the middle-high resolution simulations (blue curve) and the low-middle resolution simulations (orange curve).  The dotted curves represent the low-middle resolution scaled by different convergence factors $Q_n$, where $n$ is the order of convergence.  We find that our results are second-order convergent while the BH develops scalar hair, and become fourth-order convergent afterwards.}
    \label{fig:convergence}
\end{figure}

In Fig.~\ref{fig:convergence} we show the difference across resolutions of the quantity $\langle\phi\rangle_{\rm AH}$, which we expect to be related by the convergence factor $Q_n=(h_{\rm LR}^n - h_{\rm MR}^n)/(h_{\rm MR}^n - h_{\rm HR}^n)$~\cite{Alcubierre}, with $n$ being the order of convergence and $h_i=1/N_i$ the grid spacing for the different resolutions. We observe that the convergence order approaches four after the settlement of the gauge and the scalarisation of the BH, which is perfectly consistent with the fourth-order numerical scheme used in the code. Note that third-order interpolation is used in the AMR boundaries, causing the convergence order to be effectively between third and fourth, which reduces to almost second order as the BH develops scalar hair, which is compatible with similar convergence behaviours studied in detail in~\cite{Radia:2021smk}.

\

\section{Matrix inversion}
\label{app:inverseA}
In addition to finding a well-posed formulation, an additional challenge in evolving fully non-linear EsGB gravity is that the right-hand side of the equations of motion contains terms that depend linearly on the time derivative of the same variables being evolved. 
This means that the system we need to solve numerically is
\begin{eqnarray}\label{eq:full_system}
    \mathbb{M}\partial_tU=S~,
\end{eqnarray}
with $U$ being the evolution variables embedded in a $k$-vector (where $k$ is the number of variables), $S$ is the (also $k$-dimensional) source vector and finally $\mathbb{M}$ is a $k\times k$ matrix, which we denote as $\mathbb{M}=\mathbb{I}_k+\delta\mathbb{M}$, where $\delta\mathbb{M}$ contains linear and quadratic terms in the coupling constant $\lambda$. 

Although solving a linear system of equations may not seem like an additional challenge, several NR codes struggle with this task. This is why, in this section, to assist those NR codes, we are studying the effect of considering the full system~\eqref{eq:full_system} or a power expansion of the matrix by considering
\begin{eqnarray}
\label{eq:matrix}
    \mathbb{M}_n^{-1}=\sum\limits_{i=0}^n(-1)^i\delta\mathbb{M}^i~,
\end{eqnarray}
where we are going to set $n$ to be $1$, $2$ or $3$.

\begin{figure}
    \centering
    \includegraphics[width=1\linewidth]{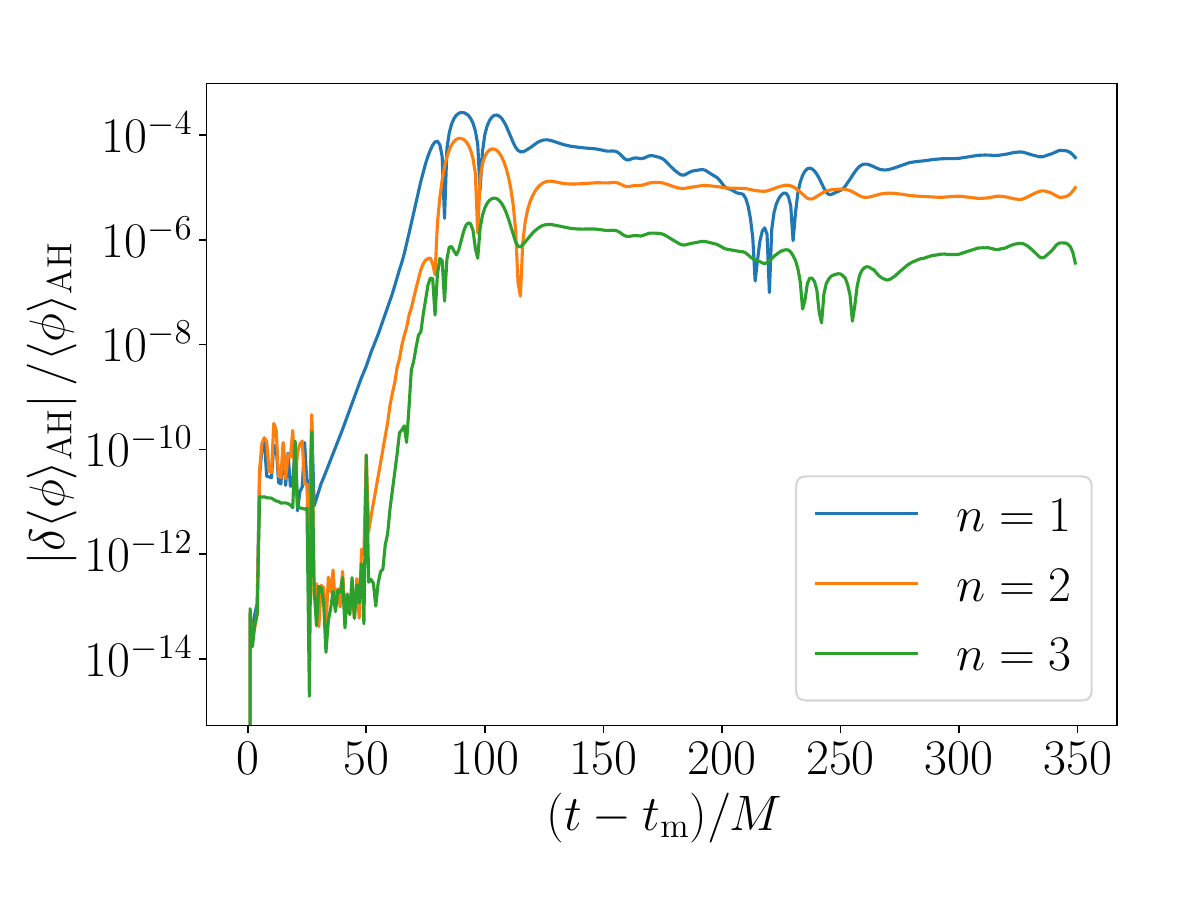}
    \caption{Relative difference across different power expansions of~\eqref{eq:matrix} of the average value of the scalar field at the apparent horizon of the remnant in a spin-induced scalarisation of a binary black hole in EsGB.}
    \label{fig:inverse}
\end{figure}

We consider the spin-induced scalarisation of a binary black hole merger studied in~\cite{AresteSalo:2023hcp}, with a quadratic-type coupling $f(\phi)=\frac{1}{2\beta}(1-e^{-\beta\phi^2})$ in the mCCZ4 formulation (see~\cite{AresteSalo:2023hcp} for further details). In Fig.~\ref{fig:inverse}, we plot the relative difference in the average value of the scalar field at the apparent horizon of the remnant between the value obtained by solving the full linear system and each one of the different power expansions of the matrix. This shows, as expected, that there is convergence to the actual solution with increasing the order $n$, though it is not clear how this would capture potential secular effects in the case of a non-GR inspiral.

\bibliography{refs.bib}
\end{document}